\documentclass[preprint,trackchanges]{aastex62}
\usepackage{amsmath}
\usepackage{natbib}
\usepackage{multirow} 

\hypersetup{linkcolor=magenta,citecolor=cyan,filecolor=yellow,urlcolor=blue}

\received{August 6, 2019}
\revised{March 3, 2020}
\accepted{March 10, 2020}
\published{Mar 11, 2020}

\shorttitle{Constraining the Hybrid Jet Model}
\shortauthors{Li.}

\begin{document}

\title{Thermal Components in Gamma-ray Bursts. II. \\Constraining the Hybrid Jet Model}

\author[0000-0002-1343-3089]{Liang Li}
\affiliation{ICRANet, Piazza della Repubblica 10, I-65122 Pescara, Italy}
\affiliation{INAF -- Osservatorio Astronomico d'Abruzzo, Via M. Maggini snc, I-64100, Teramo, Italy}
\affiliation{ICRA, Dipartimento di Fisica, Sapienza Università di Roma, P.le Aldo Moro 5, I–00185 Rome, Italy}

\correspondingauthor{Liang Li}
\email{liang.li@icranet.org}

\begin{abstract}

In explaining the physical origin of the jet composition of gamma-ray bursts (GRBs), a more general picture, i.e. the hybrid jet model (which introduced another magnetization parameter $\sigma_{0}$ on the basis of the traditional fireball model), has been well studied in Gao \& Zhang. However, it still has not yet been applied to a large GRB sample. Here, we first employ the ``top-down'' approach of Gao \& Zhang to diagnose the photosphere properties at the central engine to see how the hybrid model can account for the observed data as well, through applying a {\it Fermi} GRB sample (eight bursts) with the detected photosphere component, as presented in Li (our Paper I). We infer all physical parameters of a hybrid problem with three typical values of the radius of the jet base ($r_{0}$ = 10$^{7}$, 10$^{8}$, and 10$^{9}$ cm). We find that the dimensionless entropy for all the bursts shows $\eta\gg$ 1 while the derived (1+$\sigma_{0}$) for five bursts (GRB 081224, GRB 110721A, GRB 090719, GRB 100707, and GRB 100724) is larger than unity, indicating that in addition to a hot fireball component, another cold Poynting-flux component may also play an important role. Our analysis also shows that in a few time bins for all $r_{0}$ in GRB 081224 and GRB 110721A, the magnetization parameter at $\sim$ 10$^{15}$cm (1+$\sigma_{\rm r15}$) is greater than unity, which implies that internal-collision-induced magnetic reconnection and turbulence may be the mechanism to power the nonthermal emission, rather than internal shocks. We conclude that the majority of bursts (probably all) can be well explained by the hybrid jet problem. 

\end{abstract}


\section{Introduction} \label{sec:intro}

One of the most fundamental, yet unsolved, questions in gamma-ray burst (GRB) physics is the nature of jet composition. A crucial debate focuses on the physical origin of jet compositions---whether it is originated from a baryonic-dominated fireball \citep[e.g.,][]{2017IJMPD..2630018P} or a Poynting-flux-dominated outflow \citep[e.g.,][]{2018pgrb.book.....Z}.

An important scenario invokes a quasi-thermal component indicating a hot fireball origin, which is introduced by \cite{1986ApJ...308L..43P} and \cite{1986ApJ...308L..47G} with a pure fireball picture (composed of positron-electron pair and hot photons) in the early time. Later, it is introduced by \cite{1990ApJ...365L..55S} and \cite{1990ApJ...363..218P} with a baryon-dominated fireball framework (baryons + positron-electron pair and hot photons) in order to be consistent with the observations. In this baryon-dominated fireball scenario, the two-component spectral scenario is expected to be found in the observed spectrum during the prompt emission: a quasi-thermal component originates from the fireball photosphere \citep{1999A&A...350..334R, 2000A&A...359..855R, 2013ApJ...772...11R, 2000ApJ...530..292M, 2005ApJ...628..847R} when the optical depth goes to unity, and the emergent spectrum can be modified by the Planck-like function; a nonthermal component originates from the internal shocks \citep[IS;][]{1994ApJ...427..708P, 1994ApJ...430L..93R} in the optically thin region.

An alternative scenario invokes a nonthermal component from the synchrotron radiation of the Poynting-flux-dominated outflow \citep[e.g.,][]{2014IJMPD..2330002Z}. There are two possibilities to generate the prompt emission. One may originate from the matter-dominated emission region \citep{2002A&A...391.1141D, 2006ApJ...651..333T,2008A&A...480..305G}, while another may invoke the moderately Poynting-flux-dominated emission region via magnetic reconnection, such as an internal-collision-induced magnetic reconnection and turbulence (ICMART) event \citep{2011ApJ...726...90Z}. The GRB emergent spectrum from such a scenario is likely to be in good agreement with the observations that the typical GRB spectrum is known with the Band-like form \citep{1993ApJ...413..281B}, which is usually taken to represent a nonthermal emission component.
 
Observationally, a majority of GRBs present a nonthermal dominant Band-like spectrum. The Band function \citep{1993ApJ...413..281B} has two exponentially joined power laws, which are separated by typical peaks at $\sim$ hundreds keV, and the two power-law indices $\alpha$ (below the peak) and $\beta$ (above the peak) are typically distributed at $\sim -1.0$ and $\sim -2.2$, respectively. Alternatively, the baryon-dominated fireball scenario has been also confirmed by the observations since a quasi-thermal spectral component was found in the time-integrated or the time-resolved spectral analysis for some GRBs \cite[e.g.,][]{2004ApJ...614..827R, 2009ApJ...702.1211R,2010ApJ...709L.172R, 2019MNRAS.484.1912R, 2012MNRAS.420..468P, 2013MNRAS.433.2739I, 2015MNRAS.450.1651I, 2018MNRAS.475.1708A, 2018NatAs...2...69Z, 2019MNRAS.487.5508A, 2019ApJS..245....7L, 2019ApJS..242...16L, 2019arXiv191012615L, Ruffini-2019-p1904-4162}. These results were first discovered by the Burst And Transient Source Experiment (BATSE) on board the Compton Gamma Ray Observatory (CGRO), and later confirmed by the Gamma-ray Burst Monitor (GBM) on board the {\it Fermi} Gamma-ray Space Telescope. Meanwhile, the previous observations also revealed that thermal components exhibit diverse observational properties. They either can be detected during the entire duration of the prompt emission (e.g., GRB 100507; \citealt{2013MNRAS.432.3237G}) or may be only found at the beginning of the burst duration, and subsequently appear with a nonthermal component (e.g., \citealt{2004ApJ...614..827R} for a BATSE sample and \citealt{2019ApJS..242...16L} for a {\it Fermi} sample). On the other hand, thermal components can be grouped into two categories: the thermal-subdominant case and the thermal-dominant case. The former one invokes a thermal-subdominant component embedded into a nonthermal-dominant component \citep[e.g., 110721A;][]{2012ApJ...757L..31A}, while the later one invokes a thermal-dominant component accompanied by a nonthermal-subdominant component \citep[e.g., 090902B;][]{2010ApJ...709L.172R}  or even a `pure' blackbody (BB) emission \citep[e.g., 930214,][]{2004ApJ...614..827R}. Noteworthy, the thermal-subdominant case can account for a majority of the observations while the thermal-dominant cases are rarely observed. GRB 090902B is the most prominent one that the thermal-dominant component is observed either in the time-integrated spectral analysis (a dominant quasi-thermal component superposed on an underlying power-law component; \citealt{2009ApJ...706L.138A}) or the time-resolved spectral analysis (a multi-color BB component; \citealt{2010ApJ...709L.172R}).

A diverse spectral property found in the observations suggests that GRB ejecta may have a diverse jet composition. It may be neither fully matter-dominated ejecta nor fully magnetized outflows. More realistically, GRB outflows are likely to be a hybrid jet, which carries the two components simultaneously and launches at the central engine. In such a scenario, which component plays a leading role in the emission may be more important. Theoretically, the central engine models invoke either a hyper-accreting and fast-rotating black hole or a rapidly spinning and highly magnetized neutron star (magnetar). Therefore, a diverse jet composition is still expected: a hot component due to neutrino heating from the accretion disk or the proto neutron star, and a cold component associated with a Poynting flux launched from the black hole or the neutron star \cite[e.g.,][]{2011MNRAS.413.2031M, 2013ApJ...765..125L, 2015ApJ...801..103G}.

The hybrid jet problem reported \cite{2015ApJ...801..103G} \citep[see also][]{, 2004ApJ...614..827R} introduces another magnetization parameter $\sigma_{0}$ on the basis of the traditional fireball model, which is defined as $\sigma_{0} \equiv L_{\rm c}/L_{\rm b}$, where $L_{\rm b}$, $L_{\rm c}$, and $L_{0}=L_{\rm b}+L_{\rm c}$ are the luminosities of hot (fireball) component, cold (Poynting-flux) component, and entire wind, respectively. The rapid evolution of the photosphere emission proprieties is therefore expected to be a result of the rapid evolution of ($\eta$, $\sigma_{0}$) pairs, where $\eta$ is the dimensionless entropy of the outflow. The time-varying ($\eta$, $\sigma_{0}$) pair at the central engine could give rise to different observational characteristics. If $\eta \gg 1$ and $\sigma_{0} \ll 1$, a hot fireball with a dominant photosphere emission component could be observed (e.g., GRB 090902B). Moreover, if $\eta$ is smaller while $\sigma_{0}$ is larger, a subdominant photosphere emission component may be detected due to the thermal emission being suppressed (e.g., GRB 110721A). Finally, if $\eta$ is close to unity and $\sigma_{0} \gg 1$, we would only detect a nonthermal spectral component (e.g., GRB 080916C) since the outflow is fully dominated by a Poynting-flux component (highly magnetic outflow), and the photosphere component is completely suppressed. Therefore, the hybrid problem describes a more general picture, where the dimensionless entropy $\eta$ (hot fireball component) and the magnetization parameter $\sigma_{0}$ (cold Poynting-flux component) are two key parameters at the central engine. In such a hybrid problem, the hot matter-dominated outflow described by the pure fireball model ($\eta \gg 1$ and $\sigma_{0} \ll 1$) and the magnetized jet related to Poynting-flux-dominated outflow ($\eta \sim 1$ and $\sigma_{0} \gg 1$) are two extreme cases, which have been fully studied. However, a general picture of a hybrid system was rarely investigated before \cite{2015ApJ...801..103G}. Motivated by the introduction of the general formalism, which can cover all different possible cases, \cite{2015ApJ...801..103G} developed a theory of photosphere emission of a hybrid relativistic outflow. On the basis of an approximate dynamical evolution model of the hybrid system, two methods are proposed: the first one is the `bottom-up' approach to predict the temperature ($T_{\rm obs}$) and luminosity ($L_{\rm BB}$) of the photosphere emission for a given pair of parameters ($\eta$, $\sigma$) at central engine; the second one is the `top-down' approach to diagnose central engine parameters ($\eta$, $\sigma$) based on the observed quasi-thermal photosphere emission. They pointed out that adopting the `bottom-up' approach, we could reproduce a variety of observed GRB prompt emission spectra by {\it Fermi} for the non-dissipative photosphere model if the ($\eta$, $\sigma$) pair are allowed to vary in a wide range, and applying the `top-down' approach to GRB 110721A, we can well explain the observational data. 

Practically, it is more interesting to utilize the observational data to diagnose the properties at the central engine. Therefore, an attractive question is to know how the hybrid model can account for a large sample of {\it Fermi} bursts. Here, we address different questions based on the same {\it Fermi} GRB sample in a series of papers, focusing on the cases that the two-component spectral scenario (composited with a nonthermal component and a thermal component simultaneously) is clearly observed in their time-resolved spectral analysis. In the first paper of this series \citep[][hereafter Paper I]{2019ApJS..245....7L}, we presented the study on how the thermal components affect the nonthermal spectral parameters. In this work, we continue our systematic study by applying the same GRB sample (listed in Table 1 of Paper I) as well as the `top-down' approach of \cite{2015ApJ...801..103G} to diagnose the photosphere properties of a hybrid relativistic outflow. Meanwhile, we conduct a statistical analysis of the central engine properties of a large GRB sample. The goal in this task is to re-investigate the central engine properties by constraining the hybrid model with the observed data.

The paper is organized as follows: in Section 2, we present the methodology, which includes sample selection, data reduction, Bayesian inference, and Markov Chain Monte Carlo (MCMC) methods. In Section 3, we describe the fireball dynamical evolution of a hybrid relativistic outflow photosphere emission, and discuss some derived physical parameters. The results on constraining a hybrid jet system with our sample are presented in Section 4. The conclusions and discussions are illustrated in Section 5. Throughout the paper, the standard $\Lambda$-CDM cosmology with the parameters of $H_{0}= 67.4$ ${\rm km s^{-1}}$ ${\rm Mpc^{-1}}$, $\Omega_{M}=0.315$, and $\Omega_{\Lambda}=0.685$ is adopted \citep{2018arXiv180706209P}, and the convention $Q=10^{x}Q_{x}$ is adopted in cgs units.

\section{Methodology}

\subsection{Sample Selection and Data Reduction}

We included in our analysis all the GRBs detected by the {\it Fermi}-GBM until 2019 March 31 and having a reported photospheric component in the spectrum. We focus on the {\it Fermi}-GBM observation since it covers a broad spectral window in energy (8 keV-40 MeV), and therefore the current GRB spectral models can be fully characterized. The GBM \citep{2009ApJ...702..791M} contains 12 sodium iodide (NaI; 8keV-1MeV) detectors (n0 to n9, na and nb) as well as 2 bismuth germanate (BGO; 200keV-40MeV) detectors (b0 and b1). The Time Tagged Event (TTE) and spectral response (rsp) files are used for the selected sets of detectors. We select at most three NaI detectors in order to obtain an angle of incidence less than $60^{\circ}$ and one BGO detector with the lowest angle of incidence \citep{2012ApJS..199...19G, 2016ApJS..223...28N} for the spectral analysis. A sample of 13 {\it Fermi}-GBM such bursts are available, and the detail spectral properties of these bursts have been reviewed in paper I. The sample is presented in Table 1 of paper I. 

All temporal and spectral analysis in this work is implemented by adopting the Bayesian analysis package, i.e., the Multi-Mission Maximum Likelihood Framework (3ML, \citealt{2015arXiv150708343V}). Such a fully Bayesian approach was first applied in \cite{2019ApJS..242...16L} for a {\it Fermi}-GBM bright GRB spectral catalog (see also \citealt{2019MNRAS.490..927B, 2019ApJS..245....7L, 2019MNRAS.484.1912R,  2019ApJ...886...20Y}). The background is fitted by selecting two typical off-source (pre- and post-source) intervals with an order 0-4 polynomial for the brightest NaI detector in photon counts, and the optimal order of the polynomial is determined by a likelihood ratio test. This optimal polynomial is then applied to fit each of the 128 energy channels so as to estimate the background model for the rate in that channel. By integrating the optimal polynomial over source interval, we can obtain the background photon counts for each channel. We use the maximum likelihood-based statistics, the so-called Pgstat, given by a Poisson (observation, \citealt{1979ApJ...228..939C})-Gaussian (background) profile likelihood. Additionally, the error on the background can also be evaluated by assuming its distribution to be a Gauss. At least one background count per spectral bin is included to allow the Gaussian profile to be valid. In order to perform the time-resolved spectral analysis, we adopt the Bayesian Blocks (BBlocks; \citealt{2013ApJ...764..167S}) method with false alarm probability $p_{0}$=0.01 to rebin the TTE lightcurve of the brightest NaI detector. Subsequently, all other used detectors are binned in matching time bins. If there is more than one triggered NaI detector, we select the brightest one that has the highest significance during the source interval. Then, we utilize it for the BBlocks and background fitting. On the other hand, in order to obtain a good fitting result, we adopt $S\geq 20$ \citep[the definition of $S$ see][]{2018ApJS..236...17V} as the criterion to select the time bins that include enough source photons. This is because the spectral parameters obtained from the bins with lower $S$ values (e.g., $S<20$) typically have huge errors. To better infer physics from the spectral parameters, we selected the bursts with five $S\geq20$ time bins \citep[see also][]{2019ApJS..242...16L, 2019MNRAS.484.1912R}. Then, the sample was reduced to eight bursts with this criterion. These bursts are GRB 081224, GRB 090719, GRB 090902B, GRB 100724B, GRB 110721A, GRB 160107A, and GRB 190114C. The time-resolved spectral fitting results for each selected burst have been reported in Tables 2-9 in paper I. Please note that we take the cut-off power-law (CPL) model as a proxy for the Band model to perform the spectral analysis throughout the paper. This is because thermal components are typically observed in the left shoulder of the Band spectrum (below $E_{\rm p}$); its presence does not affect the high-energy $\beta$ index (above $E_{\rm p}$). The definition of each used model is presented in Appendix \ref{sec:model}.

\subsection{Bayesian Inference and MCMC Methods}

The parameter estimation is the primary task when performing spectral fits. Practically, we can apply either a frequentist analysis approach or Bayesian analysis method to achieve this goal. To fit a model to data, the conventional wisdom in the frequentist approach can adopt $\chi^{2}$ minimization or its variants, or more complex frequentist methods (e.g., Cstat, Pgstat) based on the Maximum Likelihood Estimation (MLE) technique. However, these traditional frequency methods are known to be problematic in some issues \citep[e.g.,][]{2010arXiv1012.3754A, 2016ApJ...827L..38G}. In current years, the Bayesian analysis technique has gained in popularity, and fitting the Bayesian statistical models by adopting MCMC methods have become a standard tool for the parameter estimation in astronomy \citep[e.g.,][]{2019MNRAS.490..927B, 2019ApJS..245....7L, 2019ApJS..242...16L}. In Bayesian inference, after the experimental data is obtained, Bayes's theorem is applied to infer and update the probability distribution of a specific set of model parameters. For instance, given an observed data set ($D$) and a profile model ($M$), the probability distribution $p(M\mid D)$, i.e., so-called $Posterior$ probability, according to the Bayes's theorem, therefore is given by 
\begin{equation}
p(M\mid D)=\frac{p(D\mid M)p(M)}{p(M)},
\end{equation}
where, $p(D\mid M$) is the likelihood that combines the model and the observed data and expresses the probability to observe (or to generate) the data set $D$ from given a model $M$ with its parameters, $p(M)$ is prior on the model parameters, and $p(D)$ is called evidence, which is constant with the purpose of normalizing.

The informative priors are adopted by using the typical spectral parameters from the $Fermi$-GBM catalogue:
\begin{eqnarray}
\label{n} \left\{ \begin{array}{ll}
\alpha_{\rm CPL} \sim \mathcal{N} (\mu=-1,\sigma=0.5) \\
E_{\rm CPL} \sim log \mathcal{N} (\mu=2,\sigma=1) & \rm keV\\
A_{\rm CPL} \sim log  \mathcal{N} (\mu=0,\sigma=2) & \rm cm^{-2}keV^{-1}s^{-1}\\
kT_{\rm BB} \sim log \mathcal{N} (\mu=2,\sigma=1) & \rm keV\\
A_{\rm BB} \sim log  \mathcal{N} (\mu=-4,\sigma=2) & \rm cm^{-2}keV^{-1}s^{-1}\\
\end{array} \right.
\label{eq:prior}
\end{eqnarray}
The posterior distribution is obtained from the prior and sampling information, and the affection from prior distribution will be weaker with the increase of the sampling information. According to the Bayes's formalism, only the simplest posterior allows for an analytic solution when we utilize Bayesian posterior sampling. However, in most cases, a high-dimensional integration is required so that the posterior is generally impossible to compute. Therefore stochastic sampling techniques, such as MCMC \citep[e.g., $emcee$;][]{2010CAMCS...5...65G} or nested \citep[e.g., MULTINEST;][]{2009MNRAS.398.1601F, 2019OJAp....2E..10F} sampling methods, are necessary to be involved. In this paper, we employ the $emcee$ to sample the posterior. For each sampling, we set the number of chains (=20), the number of learning samples (=2000) that we do not include in the final results, and the number of global samples (=10000). Since the Bayesian analysis provides the predictions described as probability distributions instead of point estimates, it provides the results that the uncertainty in the inferences could be quantified. Therefore, the parameters and error estimations can be straightforwardly obtained from the posterior distribution of any desired parameter. Probably, the posterior distribution deviates from any well-studied distributions (e.g. Gaussian or Poisson). Instead, it has a skewed and/or multi-modal form. Subsequently, the parameter estimation is obtained at A Maximum A Posteriori Probability from Bayesian posterior density distribution. The error range (or the credible level) is estimated from the Bayesian Highest Posterior Density (HPD) Interval, which covers a given percentage of the total probability density. Uncertainty therefore adopted the HPD interval at the 1$\sigma$ (68\%) Bayesian credible level, which is evaluated from the last 80\% of the MCMC 10000 samples.

\section{Derivation of the Physical Parameters of a Hybrid Problem}

GRB jets undergo different accelerate phases for different physical scenarios, as well as its acceleration laws. For the fireball model, the jet undergoes two phases: the acceleration phase and the coasting phase. In the acceleration phase, the bulk Lorentz factor $\Gamma$ would initially abide by a simply linear law with radius $r$, $\Gamma \propto r$, until reaching the saturation radius $r_{\rm s}$, where $\Gamma$ reaches to the maximum value defined by $\eta$, therefore, $\Gamma=r_{\rm s}/r_{0}\equiv \eta$. Here, $r_{0}$ is the initial size of the flow, $\eta$ is the initial internal energy per particle, which is defined as $\eta \equiv E /M c^{2}$ or $\eta \equiv L_{w}/\dot{M}c^{2}$, $\dot{M}$ is the mass injection rate, $c$ is the speed of light, and $L_{w}$ is the isotropic equivalent burst luminosity. When the photosphere radius exceeds the saturation radius (coast phase), $\Gamma \equiv \eta$ \citep{1993ApJ...405..278M, 1993MNRAS.263..861P}. Then, the flow will be in the coasting phase, and $\Gamma$ stays the same at the maximum value until it gets to the IS radius. Finally, it enters into the deceleration phase.

For the Poynting-flux-dominated outflow, the magnetized jet may encounter three phases: the rapid acceleration phase, the slow acceleration phase, and the coasting phase. Two acceleration phases have different acceleration laws, which are separated by the `Magneto-Sonic point' at  $r_{\rm ra}$ (the radius of rapid acceleration).  The acceleration law may be described with a power-law scaling, $\Gamma\propto r^{\lambda}$, with power-law index ranging within $\frac{1}{2}<\lambda\leq 1$ \cite[e.g.,][]{2009MNRAS.394.1182K,2011MNRAS.411.1323G} during the rapid acceleration phase ($r_{0}<r<r_{\rm ra}$), while it may be written as a general scaling, $\Gamma\propto r^{\delta}$, with $0<\delta \leq \frac{1}{3}$ \cite[e.g.,][]{2011ApJ...733L..40M, 2012ApJ...755...12V} during the slow acceleration ($r_{\rm ra}<r<r_{\rm s}$), until reaching the coasting radius $r_{\rm c}$ where $\Gamma$ reaches $\sigma_{0}$. Finally, the flow will be in the coasting phase ($r>r_{\rm s}$). In this phase, one has $\Gamma \equiv \Gamma_{\rm c}$.

For the hybrid jet system, the jet dynamic still undergoes three phases separated by $r_{\rm ra}$ and $r_{\rm s}$. Initially, it is the rapid acceleration phase dominated by the thermal acceleration ($r_{0}<r<r_{\rm ra}$) until the rapid acceleration radius $r_{\rm ra}$, then the slow acceleration phase dominated by the magnetic acceleration ($r_{\rm ra}<r<r_{\rm s}$) until the saturation radius $r_{\rm s}$ \citep{1997ApJ...482L..29M, 2003ApJ...596.1104V, 2015ApJ...801..103G}; finally, it is the coasting phase ($r>r_{\rm s}$), where $r_{\rm ra}$ it is defined by the larger one of the thermal coasting radius or the magneto-sonic point. Therefore, the acceleration law can approximately be written as $\Gamma\propto r$ for the rapid acceleration phase, $\Gamma\propto r^{\delta}$ during the slow acceleration phase, and $\Gamma \equiv \Gamma_{\rm c}$ when in the coasting phase.

In this paper, we focus on applying the observed data to diagnose the properties at the central engine for a hybrid problem. Such an interesting question was first worked out by \cite{2007ApJ...664L...1P} for the pure fireball model. Three observed quantities can be obtained from the spectral fits: the observed BB temperature $kT_{\rm obs}$, the observed BB flux $F_{\rm BB}$, and the observed total flux $F_{\rm obs}$ (thermal+nonthermal). Once we know all of these three observed quantities ($kT_{\rm obs}$, $F_{\rm BB}$, and $F_{\rm obs}$), we can infer the values of the isotropic equivalent luminosity of the thermal component $L_{\rm BB}$, the Lorentz factor of the bulk motion of the flow at the photospheric radius $\eta$, and the physical size at the base of the flow $r_{0}$, through applying the method developed in \cite{2007ApJ...664L...1P} for the case\footnote{Note that the method cannot be applied for the cases of $r_{\rm ph}<r_{\rm c}$ since due to degeneracy.} of $r_{\rm ph}>r_{\rm c}$. In the pure fireball model, three unknowns ($L_{\rm BB}$, $\eta$, and $r_{0}$) can be solved by three known observed parameters ($kT_{\rm obs}$, $F_{\rm BB}$, and $F_{\rm obs}$).

In the hybrid problem, there are four unknown parameters at the central engine ($L_{\rm w}$, $r_{0}$, $\eta$, and $\sigma_{0}$) since another parameter $\sigma_{0}$ is introduced. Hence, it is unlikely to solve all of these four unknown parameters from the observed data. In this scenario, considering a realistic central engine, \cite{2015ApJ...801..103G} suggested that assuming a constant $r_{0}$ throughout a burst for analysis could be more reasonable. Following this concept, we can also derive all the relevant photosphere properties for a hybrid problem (e.g., $\eta$, 1+$\sigma_{0}$, $r_{\rm ph}$, $\Gamma_{\rm ph}$, 1+$\sigma_{\rm ph}$, 1+$\sigma_{\rm r15}$), where $r_{\rm ph}$ is the photosphere radius, $\Gamma_{\rm ph}$ is the bulk Lorentz factor at $r_{\rm ph}$, 1+$\sigma_{\rm ph}$ is the magnetization parameter at $r_{\rm ph}$, and 1+$\sigma_{\rm r15}$ is the magnetization parameter at $10^{15}$ cm. Since the BB component is predicted only in the non-dissipative photosphere models, we pay special attention to such models. On the other hand, the magnetically dissipative photosphere models predict a much higher $E_{\rm p}$, which is disfavored by the observed spectrum \citep{2015ApJ...802..134B}. There are six different regimes for the photosphere properties in the hybrid system\footnote{This is because the photosphere radius $r_{\rm ph}$ can be in three different regimes separated by $r_{\rm ra}$ and  $r_{\rm c}$, and the Lorentz factor at $r_{\rm ra}$, $\Gamma_{\rm ra}$ has two different possible values for different central engine parameters: $\eta>(1+\sigma)^{1/2}$ and $\eta<(1+\sigma)^{1/2}$.}, which can be applied for outflows in the case of both sub-photospheric magnetic dissipation and non sub-photospheric magnetic dissipation. Regime I: $\eta>(1+\sigma)^{1/2}$ and $r_{\rm ph}<r_{\rm ra}$; Regime II: $\eta>(1+\sigma)^{1/2}$ and $r_{\rm ra}<r_{\rm ph}<r_{\rm c}$; Regime III: $\eta>(1+\sigma)^{1/2}$ and $r_{\rm ph}>r_{\rm c}$; Regime IV: $\eta<(1+\sigma)^{1/2}$ and $r_{\rm ph}<r_{\rm ra}$; Regime V: $\eta<(1+\sigma)^{1/2}$ and $r_{\rm ra}<r_{\rm ph}<r_{\rm c}$; Regime VI: $\eta<(1+\sigma)^{1/2}$ and $r_{\rm ph}>r_{\rm c}$. Similarly, the central engine parameters cannot be inferred in the case of $r_{\rm ph}<r_{\rm ra}$ due to the degeneracy for the hybrid problem (regimes I and IV). Therefore, our analysis will focus on the case of  $r_{\rm ph}>r_{\rm ra}$ (regimes II, III, V, and VI).

\section{Results}

\begin{figure*}
\includegraphics[angle=0, scale=0.40]{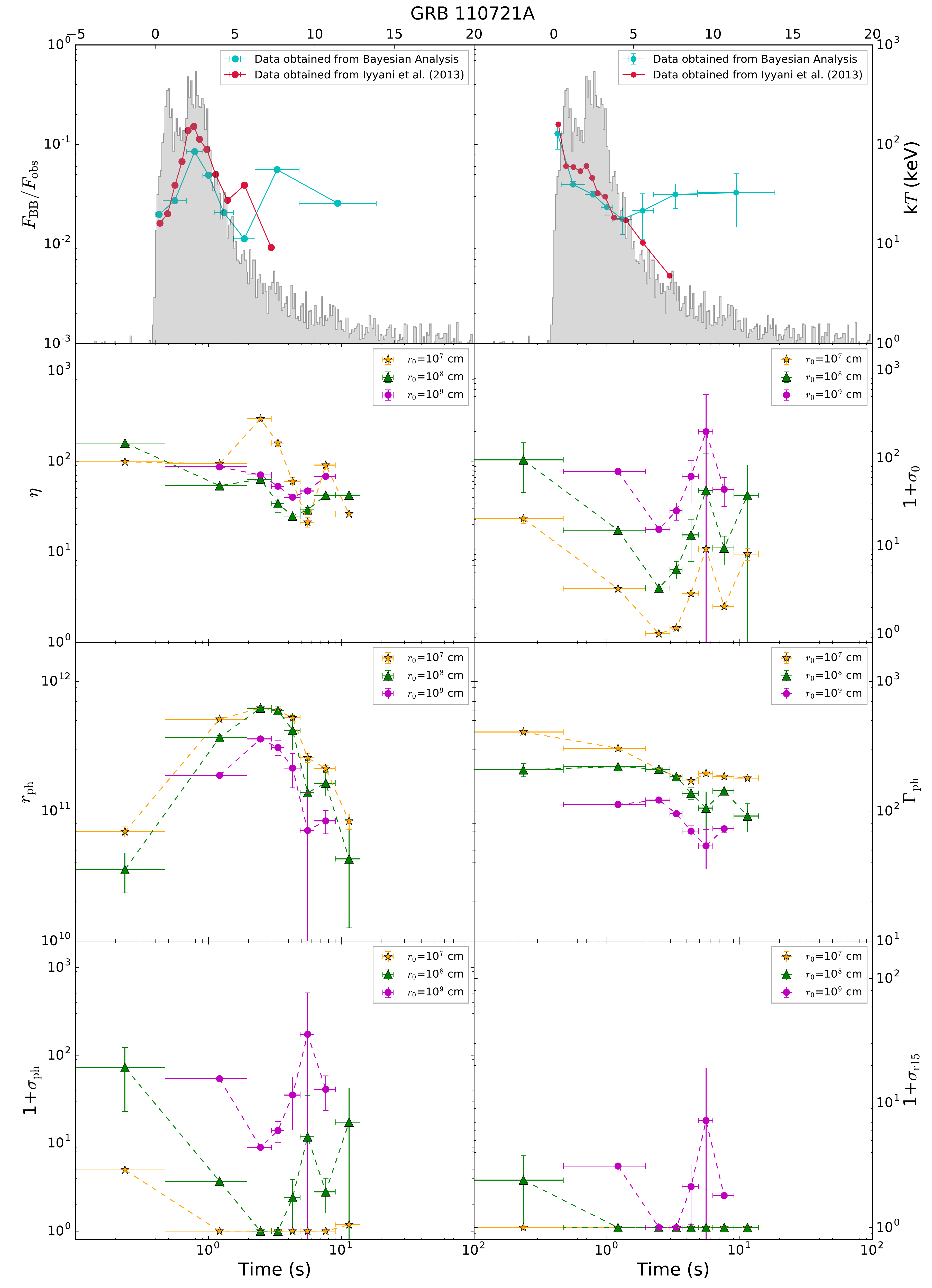}\centering
\caption{Temporal evolution of thermal flux ratio, BB temperature, and all physical parameters ($\eta$, 1+$\sigma_{0}$, $r_{\rm ph}$, $\Gamma_{\rm ph}$, 1+$\sigma_{\rm ph}$, 1+$\sigma_{\rm r15}$) of the hybrid problem for GRB 110721A. The fitted parameters are obtained from the best fitting of the CPL+BB model by using Bayesian analysis + MCMC method. The physical parameters are calculated by using top-down approach of \cite{2015ApJ...801..103G}, and considering the case in a non-dissipative photosphere. Regime judgment is used from Table 2 of \cite{2015ApJ...801..103G}. The redshift of $z$=0.382 is adopted. Three values of $r_{0}$ are used and different colors represent different values of $r_{0}$: $r_{0}$=10$^{7}$ cm (orange), $r_{0}$=10$^{8}$ cm (green), and $r_{0}$=10$^{9}$ cm (purple). Note that the two observed parameters (top panels) share the same time scale in the linear-log plots while the physical parameters (all the other panels) share the same time scale in the log-log plots.}\label{110721200}
\end{figure*}

\begin{figure*}
\includegraphics[angle=0, scale=0.40]{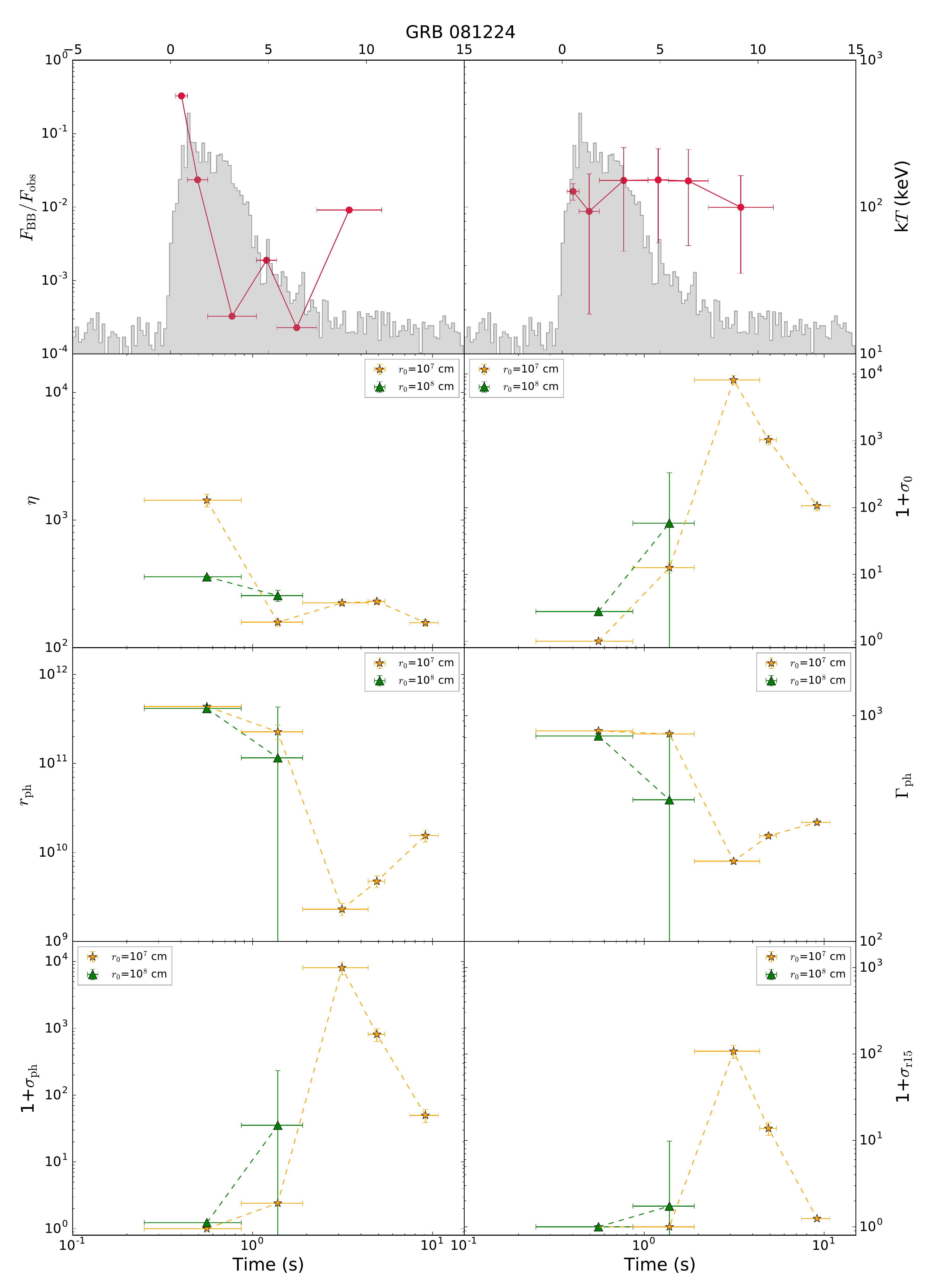}\centering
\caption{Same as Figure \ref{110721200}, but for GRB 081224.}\label{081224887}
\end{figure*}

\begin{figure*}
\includegraphics[angle=0, scale=0.40]{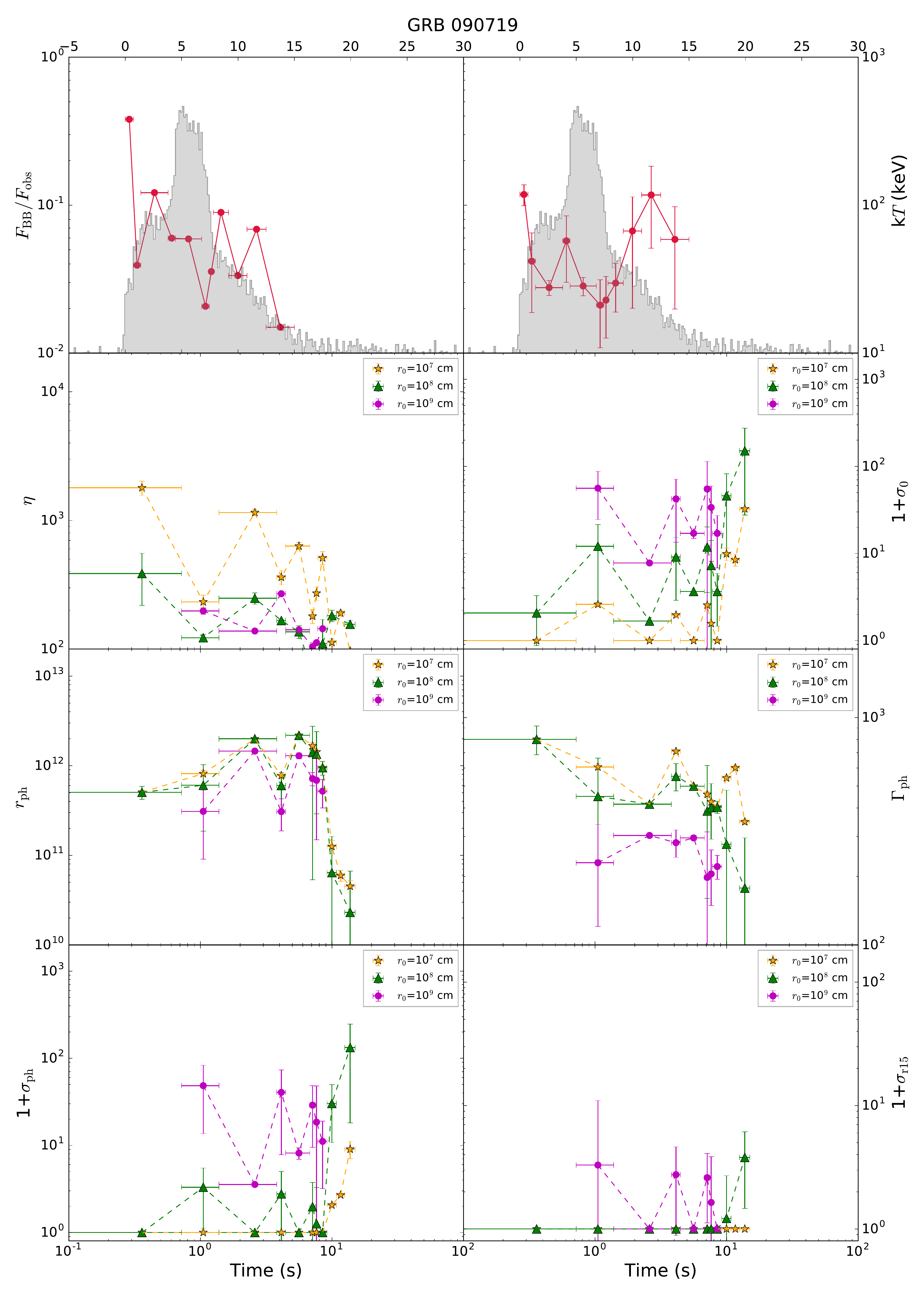}\centering
\caption{Same as Figure \ref{110721200}, but for GRB 090719.}\label{090719063}
\end{figure*}

\begin{figure*}
\includegraphics[angle=0, scale=0.40]{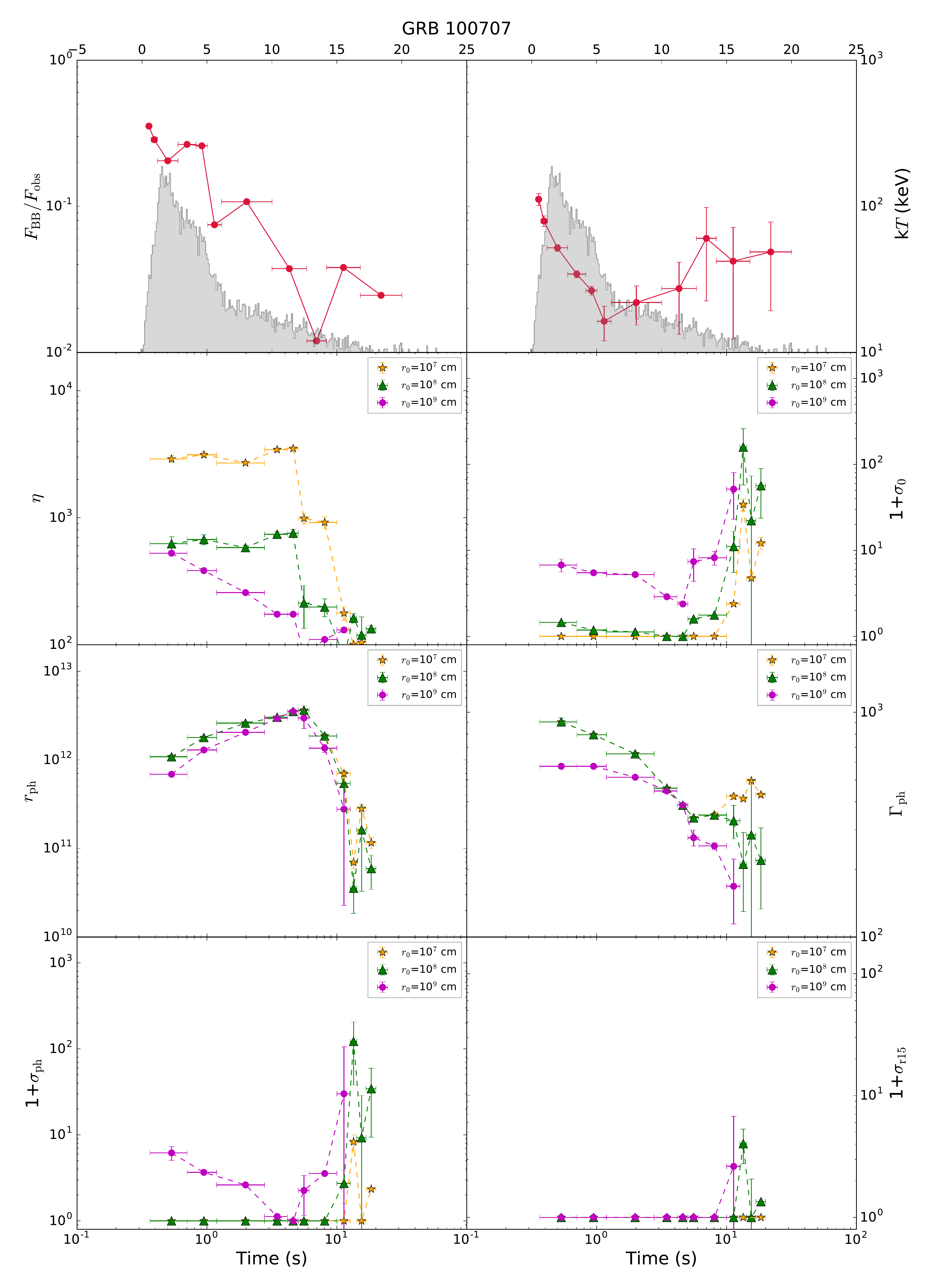}\centering
\caption{Same as Figure \ref{110721200}, but for GRB 100707.}\label{100707032}
\end{figure*}

\begin{figure*}
\includegraphics[angle=0, scale=0.40]{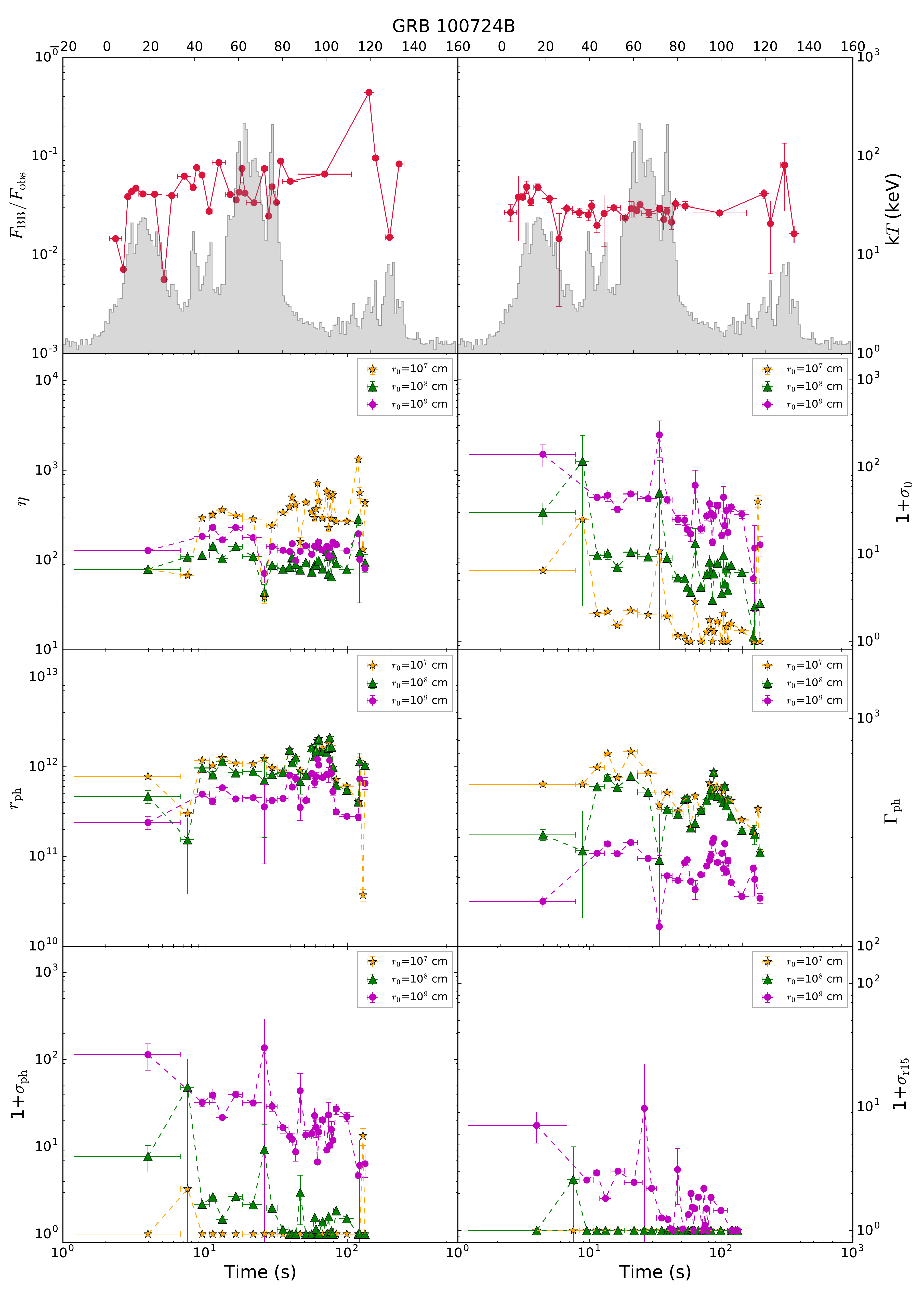}\centering
\caption{Same as Figure \ref{110721200}, but for GRB 100724B.}\label{100724029}
\end{figure*}

\begin{figure*}
\includegraphics[angle=0, scale=0.40]{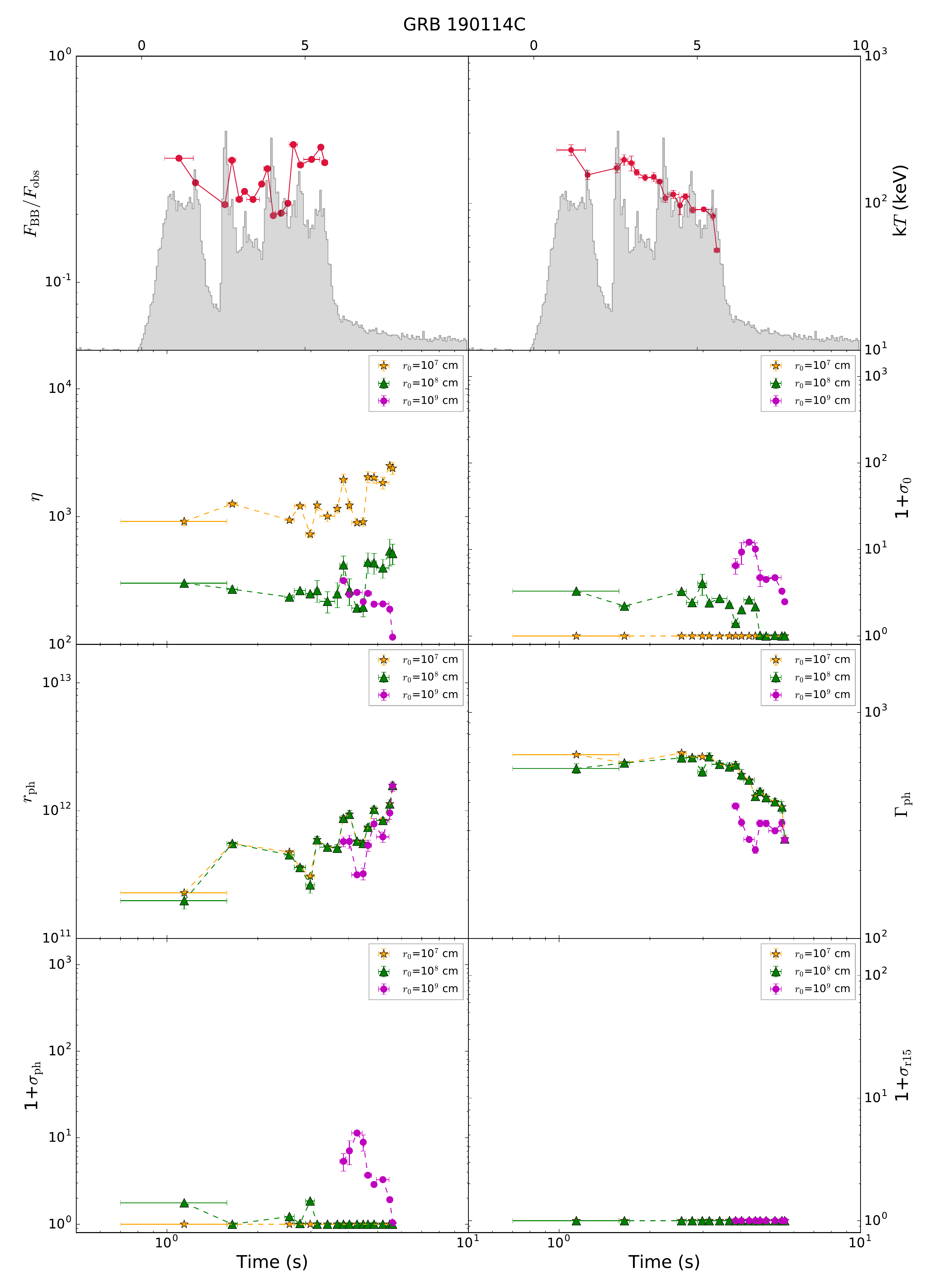}\centering
\caption{Same as Figure \ref{110721200}, but for GRB 190114C.}\label{190114873}
\end{figure*}

\begin{figure*}
\includegraphics[angle=0, scale=0.40]{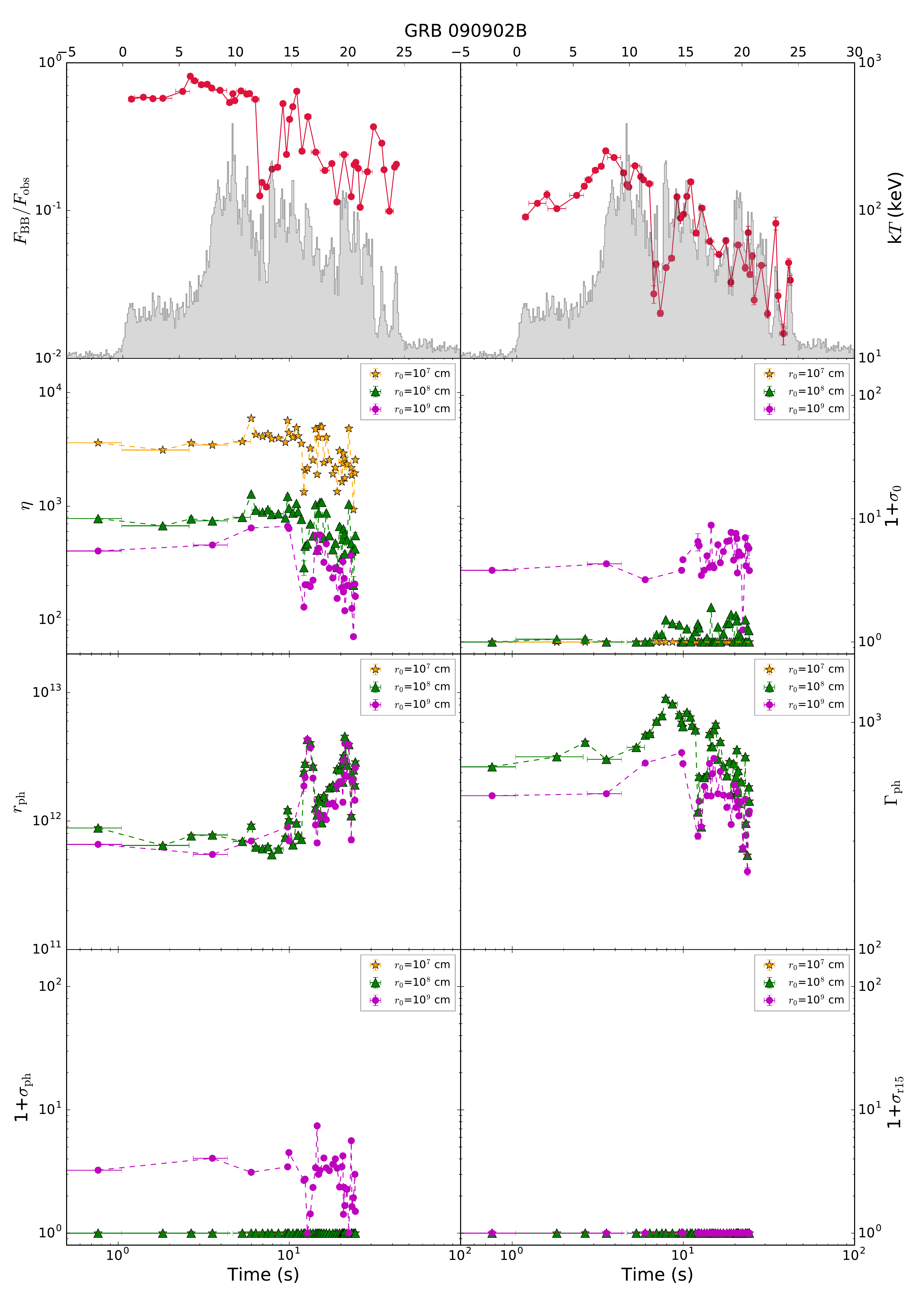}\centering
\caption{Same as Figure \ref{110721200}, but for GRB 090902B.}\label{090902462}
\end{figure*}

\begin{figure*}
\includegraphics[angle=0, scale=0.40]{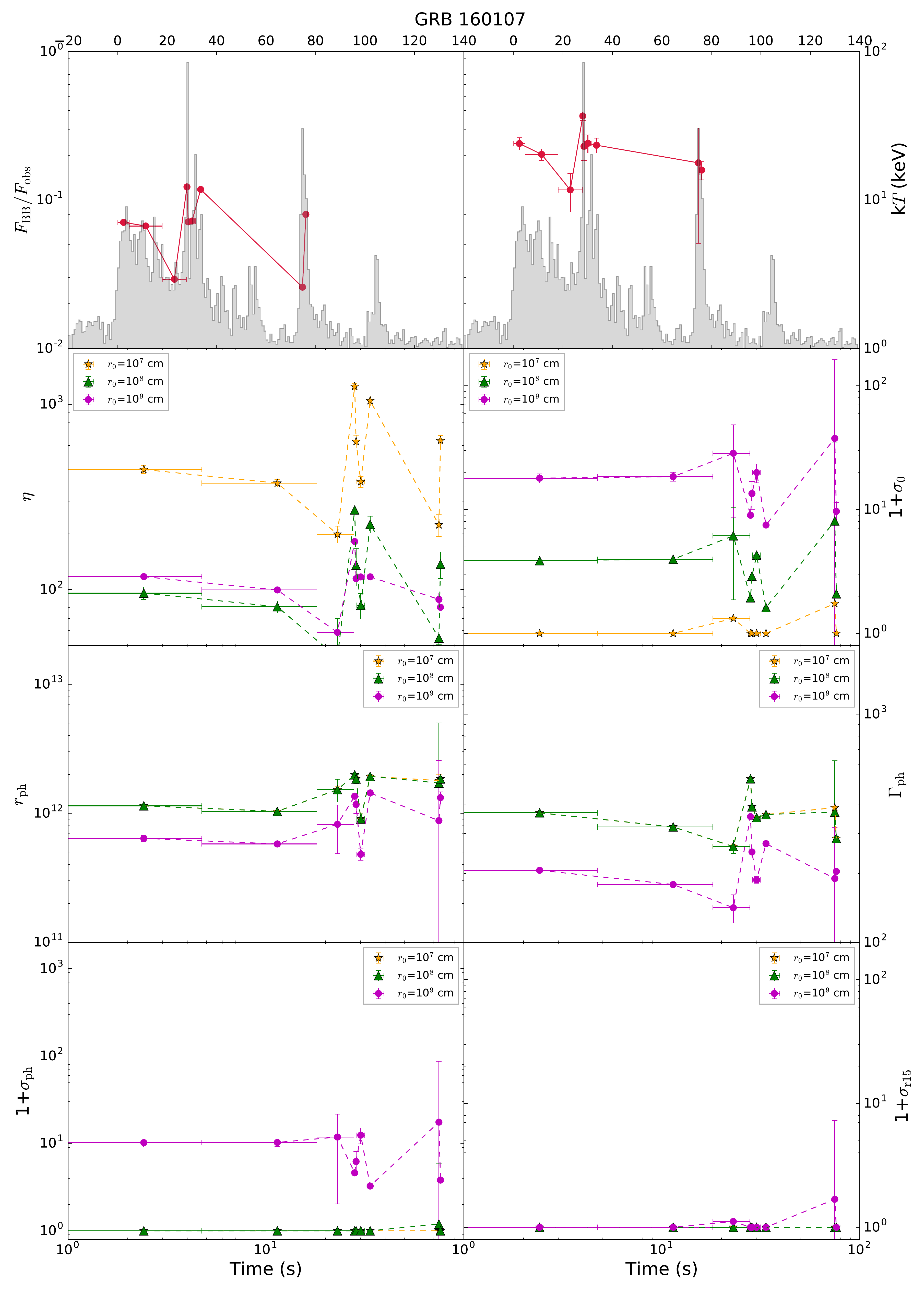}\centering
\caption{Same as Figure \ref{110721200}, but for GRB 160107A.}\label{160107931}
\end{figure*}

We report the properties of the physical parameters of our sample for the hybrid problem in Table \ref{tab:1}. For each burst, we present the results with three different $r_{0}$ values: $r_{0}$=10$^{7}$ cm, $r_{0}$=10$^{8}$ cm, and $r_{0}$=10$^{9}$ cm. For the bursts without redshift, we utilize a typical value ($z=2$) instead. We will discuss further in \S \ref{Conclusion} for why these values are adopted. By using the `top-down' approach of \cite{2015ApJ...801..103G}, we then derive all the relevant parameters of the hybrid problem at the central engine ($\eta$, 1+$\sigma_{0}$, $r_{\rm ph}$, $\Gamma_{\rm ph}$, 1+$\sigma_{\rm ph}$, 1+$\sigma_{\rm r15}$). The inferred physical parameters depend on an assumed constant $r_{0}$ for the hybrid problem, which has been suggested in \cite{2015ApJ...801..103G}. We present the temporal properties of these physical parameters as well as the two observed parameters (the ratio between the BB to total flux $F_{\rm BB}$/$F_{\rm obs}$ and the BB temperature $T_{\rm obs}$) for each burst in Figures \ref{110721200}-\ref{160107931}.

We find that the temporal properties vary from burst to burst, even the same burst uses different values of $r_{0}$. To better express the temporal evolution properties of physical parameters, we denote different types (see Table \ref{tab:1} and below for detail definitions). Different temporal properties of the physical parameters may imply different central engine properties. For instance, (1+$\sigma_{0}$) is expected to initially increase with time in some engine models \citep[e.g.,][]{2011MNRAS.413.2031M}. The pure fireball model predicts $\Gamma_{\rm ph}$ initially rises with time, whereas both IS and ICMART scenarios expect $\Gamma_{\rm ph}$ decreases with time.

The analysis of characteristics on the temporal evolution of physical parameters has led up to identifying the following unique features of our sample:
\begin{itemize}
\item GRB 110721A. The time-resolved spectral analysis shows that 10 time bins that satisfy with our selection criteria (see \S 2). Through the regime judgment, we obtain 8, 8, and 6 time bins for $r_{0}$=10$^{7}$ cm, $r_{0}$=10$^{8}$ cm, and $r_{0}$=10$^{9}$ cm, respectively. We therefore use these time bins for physical inference. The redshift $z$=0.382 is reported in \cite{2011GCN.12193....1B}. Figure \ref{110721200} presents the temporal evolution of physical parameters with different $r_{0}$ values for the hybrid problem. Throughout the burst duration, we find that for all $r_{0}$, the derived $\eta \gg$ 1 for all time bins and the derived (1+$\sigma_{0}$) is greater than unity for a majority of time bins. The results indicate that in addition to a hot fireball component, another cold Poynting-flux component may also play an important role at the central engine. Moreover, we find that $\eta$ shows a monotonic {\it decreasing} ({\it d.}) trend while (1+$\sigma_{0}$) exhibits a {\it decrease}-to-{\it increase} ({\it d.}-to-{\it i.}) trend, which is consistent with what is expected in some engine models \citep[e.g.,][]{2011MNRAS.413.2031M}. On the other hand, $r_{\rm ph}$ presents an {\it increase-to-decrease} ({\it i.}-to-{\it d.}) trend and $\Gamma_{\rm ph}$ also shows a monotonic {\it decrease}. Interestingly, a good fraction of time bins for both (1+$\sigma_{\rm ph}$) and (1+$\sigma_{\rm r15}$) are above unity, which suggests that the radiation mechanism of nonthermal components for this burst may be an ICMRAT event rather than IS. 

The fitted parameters (e.g., $F_{\rm BB}$, $F_{\rm obs}$, and $kT$) obtained from different analysis (frequency or Bayesian) methods may differ. The inferred physical parameters by utilizing the fitted parameters obtained from \cite{2013MNRAS.433.2739I} are shown in Figure \ref{110721200A1} (z=0.382) and \ref{110721200A2} (z=3.512), while those from our Bayesian analysis are presented in Figure \ref{110721200}.

\item GRB 081224. \cite{2014ApJ...784...17B} reported the time-resolved spectral analysis. They suggested that the acceptable spectral fits required an additional BB component to the synchrotron component. This result is confirmed by our Bayesian analysis. No redshift is reported for the burst, we therefore use a typical value ($z$=2). In total, we include five time bins for the analysis. The numbers of time bins of the regime judgment for each value of $r_{0}$ are listed in Table \ref{tab:1} (Column 4). We find that $\eta \gg$ 1 for all time bins, and they show a monotonic {\it decreasing} behavior (Figure \ref{081224887}). While (1+$\sigma_{\rm 0}$) rapidly {\it rise} from $\sim$ 1 then {\it decay} later as a power law. When it reaches its maximum value, $\eta$ reaches its minimum value correspondingly. The results suggest that the outflow for the burst could be dominated by a cold Poynting-flux component. On the other hand, both $r_{\rm ph}$ and $\Gamma_{\rm ph}$ present a {\it decrease-to-increase} trend. A few time bins for both (1+$\sigma_{\rm ph}$) and (1+$\sigma_{\rm r15}$) are above unity, which indicates that the ICMRAT event may be the preferred model than IS to explain the nonthermal component for the burst.

\item GRB 090719. The burst was also revealed that the best model for the spectral fits require an additional BB component \citep{2014ApJ...784...17B}, which is consistent with our Bayesian analysis. We obtain 12 time bins. Among these, 11, 10, and 7 bins satisfy with the regime judgment for $r_{0}$=10$^{7}$ cm, $r_{0}$=10$^{8}$ cm, and $r_{0}$=10$^{9}$ cm, respectively. Again, we apply a value of $z$=2 as a proxy for redshift. The temporal evolution of the physical parameters and the observational parameters are shown in Figure \ref{090719063}. Still, we find that $\eta \gg$ 1 for all time bins with moderate-$\sigma_{0}$ for most time bins, i.e., (1+$\sigma_{0})$$>$1. Except that, we also find that the derived (1+$\sigma_{0}$) shows monotonic {\it increases} ({\it i.}) with time, which is consistent with the expectation in some central engine models \citep[e.g.,][]{2011MNRAS.413.2031M}. On the other hand, $r_{\rm ph}$ and $\Gamma_{\rm ph}$ generally present a {\it flat-to-decrease} ({\it f.}-to-{\it d.}) trend. A few time bins for (1+$\sigma_{\rm ph}$) as well as (1+$\sigma_{\rm r15}$) are slightly greater than unity while the others are close to unity. One can tentatively draw the conclusion that a strongly cold Poynting-flux component is found in this burst. It is not clear that whether the ICMRAT event or IS is the radiation mechanism of nonthermal components for the burst, because it depends on which $r_{0}$ value is the true size at the central engine.  

\item GRB 100707. The burst was also analyzed in \cite{2014ApJ...784...17B}, and it was also suggested that an additional thermal component should be added to the spectral fitting in order to obtain an acceptable fitting. We include 11 time bins, and 11, 11, and 8 time bins satisfy with the regime judgment for $r_{0}$=10$^{7}$ cm, $r_{0}$=10$^{8}$ cm, and $r_{0}$=10$^{9}$ cm, respectively. Redshift is still adopted a typical value, namely, $z$=2. All time bins show $\eta \gg$ 1, and present {\it flat-to-decrease} evolution, while (1+$\sigma_{0}$) is initially close to unity and then {\it increases} ({\it f.}-to-{\it i.}) rapidly (Figure \ref{100707032}). Moreover, $r_{\rm ph}$ show an {\it increase-to-decrease} temporal trend while $\Gamma_{\rm ph}$ generally present a slow-to-fast {\it decrease}. We find that (1+$\sigma_{\rm ph}$) shows a very similar behavior in contrast to (1+$\sigma_{\rm 0}$). Almost all of time bins for (1+$\sigma_{\rm r15}$) are close to unity (see Figure \ref{100707032}), implying that IS plays a more important role than ICMRAT to explain the nonthermal emission. The results suggest that a cold Poynting-flux component plays a prominent role at a later time since the derived (1+$\sigma_{0}$) is larger than unity for all $r_{0}$. This is consistent with the observation that the thermal flux ratio ($F_{\rm BB}/F_{\rm obs}$) presents a strong temporal evolution (it decays rapidly with time) during the duration \citep{2019ApJS..245....7L}.

\item GRB 100724B. After conducting the detailed time-resolved spectral analysis, \cite{2011ApJ...727L..33G} pointed out that the spectrum of GRB 100724B is dominated by the typical Band function, including a statistically significant thermal contribution. This burst is very bright, and many more time bins are available for the analysis. In total, we obtained 33 time bins. There are 32, 31, and 30 time bins respectively conform to the regime judgment for $r_{0}$=10$^{7}$ cm, $r_{0}$=10$^{8}$ cm, and $r_{0}$=10$^{9}$ cm. There is no redshift observation, and we still utilize $z$= 2. Thermal flux ratio slightly {\it increases} while BB temperatures generally show a {\it flat} ({\it f.}) trend (Figure \ref{100724029}). We find that $\eta \gg$ 1 for all time bins and shows a slightly monotonic-{\it increase} trend, while the derived (1+$\sigma_{0}$) $>$1 for almost all of time bins and presents a monotonic {\it decrease} trend. On the other hand, $r_{\rm ph}$ and $\Gamma_{\rm ph}$ generally present a {\it flat} behavior. We also find that the derived (1+$\sigma_{\rm ph}$) and (1+$\sigma_{\rm r15}$) show a $r_{0}$-dependent behavior, i.e., nearly all time bins for a large $r_{0}$ ($r_{0}$=10$^{9}$cm) are beyond unity but for a smaller $r_{0}$ ($r_{0}$=10$^{7}$ cm) are close to unity. This implies that whether ICMRAT or IS is the mechanism to power the nonthermal emission depends on which $r_{0}$ is the true size at the central engine. A firm conclusion that can be drawn for this burst is that a prominent Poynting-flux component and a fireball component are both observed. Therefore, a hybrid jet problem should be considered. 

\item GRB 190114C. After performing the detail time-resolved spectral analysis and model comparisons, \cite{2019arXiv190107505W} recently reported that during the first spike of the burst, adding a BB greatly improves the fitting over the CPL model---around 2.7 to $\sim$ 5.5 s. This burst is a very bright, and we include 18 time bins from 2.7 to 5.5 s. Through regime judgment, 17, 17, and 9 time bins are obtained for $r_{0}$=10$^{7}$ cm, $r_{0}$=10$^{8}$ cm, and $r_{0}$=10$^{9}$ cm, respectively. Redshift is adopted $z$= 0.424 reported by \cite{2019GCN.23695....1S}. The thermal flux ratio is very high (Figure \ref{190114873}) and without significant evolution, with an average $\sim$ 30\% for all time bins, which is much higher than the typical observations. BB temperature generally shows a monotonic {\it decreasing} behavior. All time bins show $\eta \gg$ 1 while for a majority of time bins the derived (1+$\sigma_{0}$) is $\sim$ unity for a small $r_{0}$ (10$^{7}$cm) but above unity for a large $r_{0}$ (10$^{9}$ cm) ($r_{0}$-dependent). Also, we find that $r_{\rm ph}$ shows {\it increases} while $\Gamma_{\rm ph}$ generally present a {\it flat}-to-{\it decrease} temporal trend. Interestingly, no time bins for all $r_{0}$ the derived (1+$\sigma_{\rm r15}$) are above unity ($r_{0}$-independent), while (1+$\sigma_{\rm ph}$) show a $r_{0}$-dependent behavior. 

\item GRB 090902B. The burst shows the thermal dominate form \citep{2010ApJ...709L.172R}, both in the time-integrated or time-resolved spectral analysis. Moreover, GRB 090902B is a very bright burst, and 48 time bins are obtained. The redshift of $z$=1.822 was measured by \cite{2009GCN..9873....1C}. Three radii ($r_{0}$=10$^{7}$ cm, $r_{0}$=10$^{8}$ cm, and $r_{0}$=10$^{9}$ cm) correspond to 47, 47, 33 bins, satisfying with the regime judgments. In Figure \ref{090902462}, we present temporal evolutions of all physical parameters, ratios, and BB temperature. The thermal flux ratios reach a very high value at early times, with an average value $\sim$ 70\%, then decrease to $\sim$ 20 \% at later times. We find that all time bins show $\eta \gg$ 1 and show dramatic {\it flat}-to-{\it decrease} properties. Both (1+$\sigma_{0}$) and (1+$\sigma_{r15}$) exhibit a $r_{0}$-dependent behavior, while there is no time bin for all $r_{\rm 0}$ the derived (1+$\sigma_{\rm r15}$) is greater than unity, indicating that IS is the mechanism to power the nonthermal emission. Interestingly, $r_{\rm ph}$ shows a {\it flat-to-increase} behavior whereas $\Gamma_{\rm ph}$ presents a {\it flat-to-decrease} behavior.

\item GRB 160107A. The burst is another case which reveals the thermal-dominant form, and \cite{2018PASJ...70....6K} suggests the best spectral model is PL+BB. No redshift is reported and $z=$2 is still adopted. All nine time bins still show $\eta \gg$ 1 and showing a {\it flat} behavior (Figure \ref{160107931}). Furthermore, we find that both (1+$\sigma_{0}$) and (1+$\sigma_{\rm ph}$), as well as $r_{\rm ph}$ and $\Gamma_{\rm ph}$, also present a {\it flat} temporal trend throughout the duration. $r_{0}$-dependent behavior is significantly found in both (1+$\sigma_{0}$) and (1+$\sigma_{\rm ph}$). We do not find any time bin where (1+$\sigma_{\rm r15}$) is greater than unity, implying that IS is the dominant mechanism to power the nonthermal emission. 
\end{itemize}

\begin{figure*}
\includegraphics[angle=0, scale=0.45]{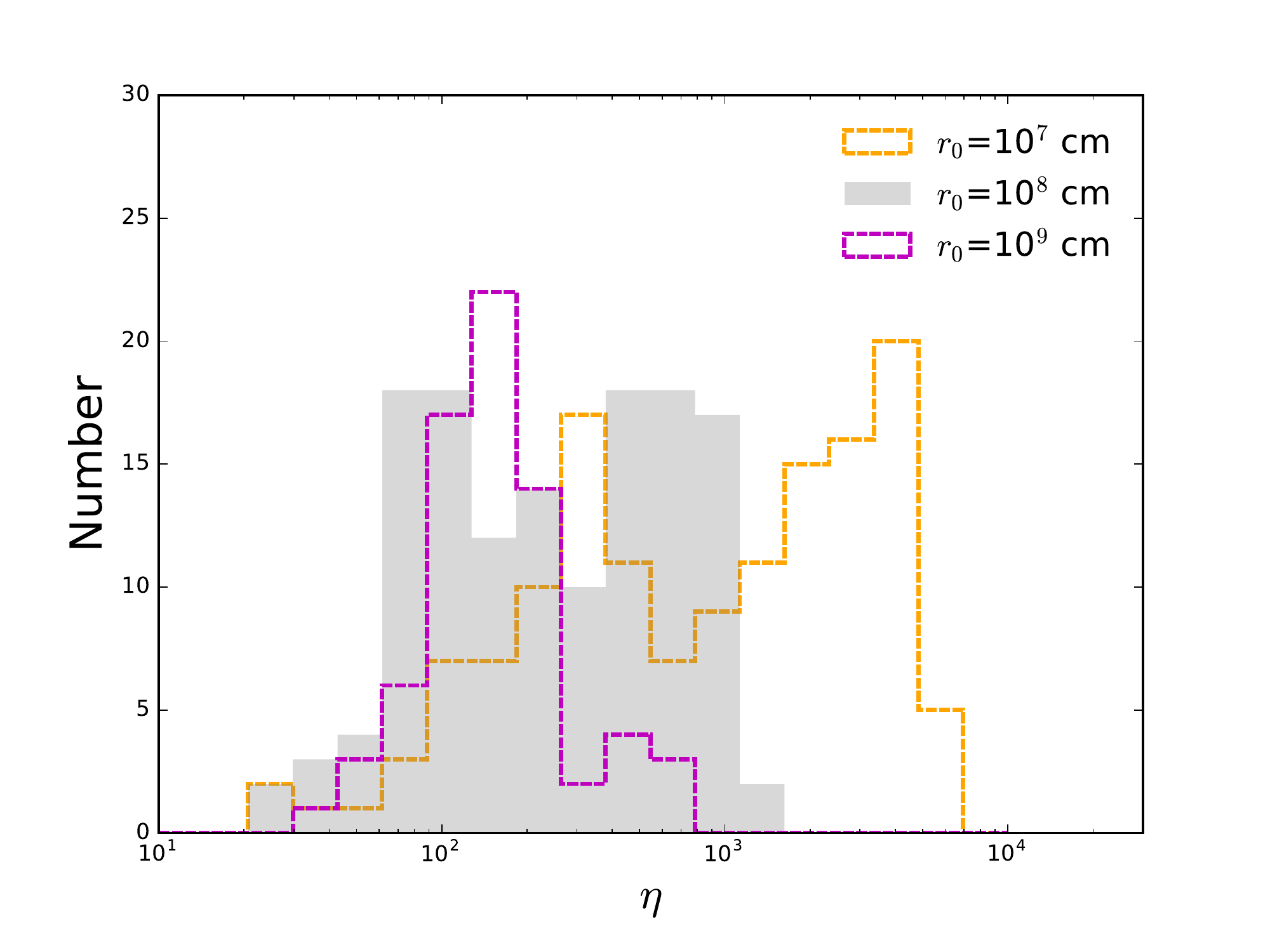}\centering
\includegraphics[angle=0, scale=0.45]{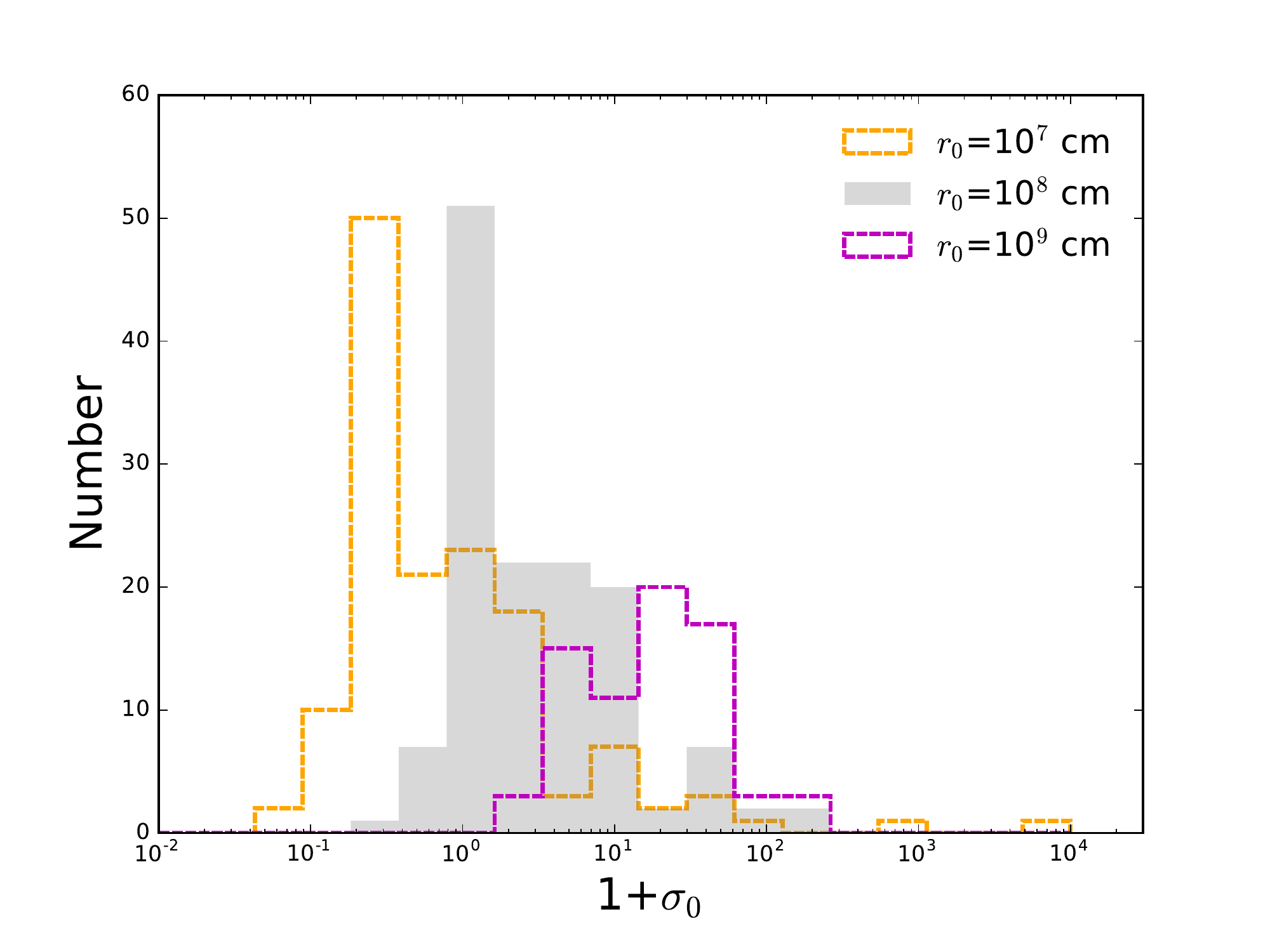}\centering\\
\includegraphics[angle=0, scale=0.45]{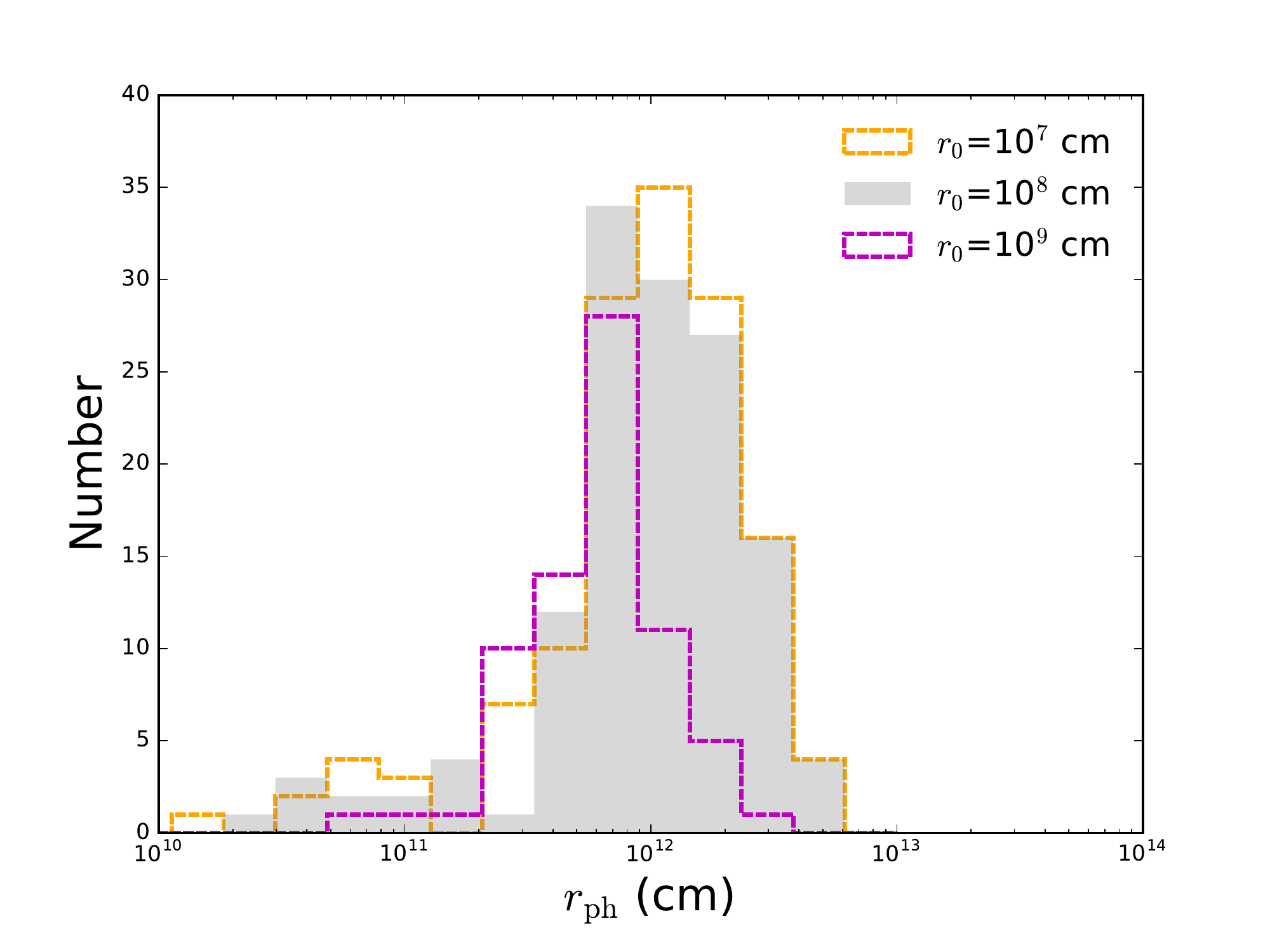}\centering
\includegraphics[angle=0, scale=0.45]{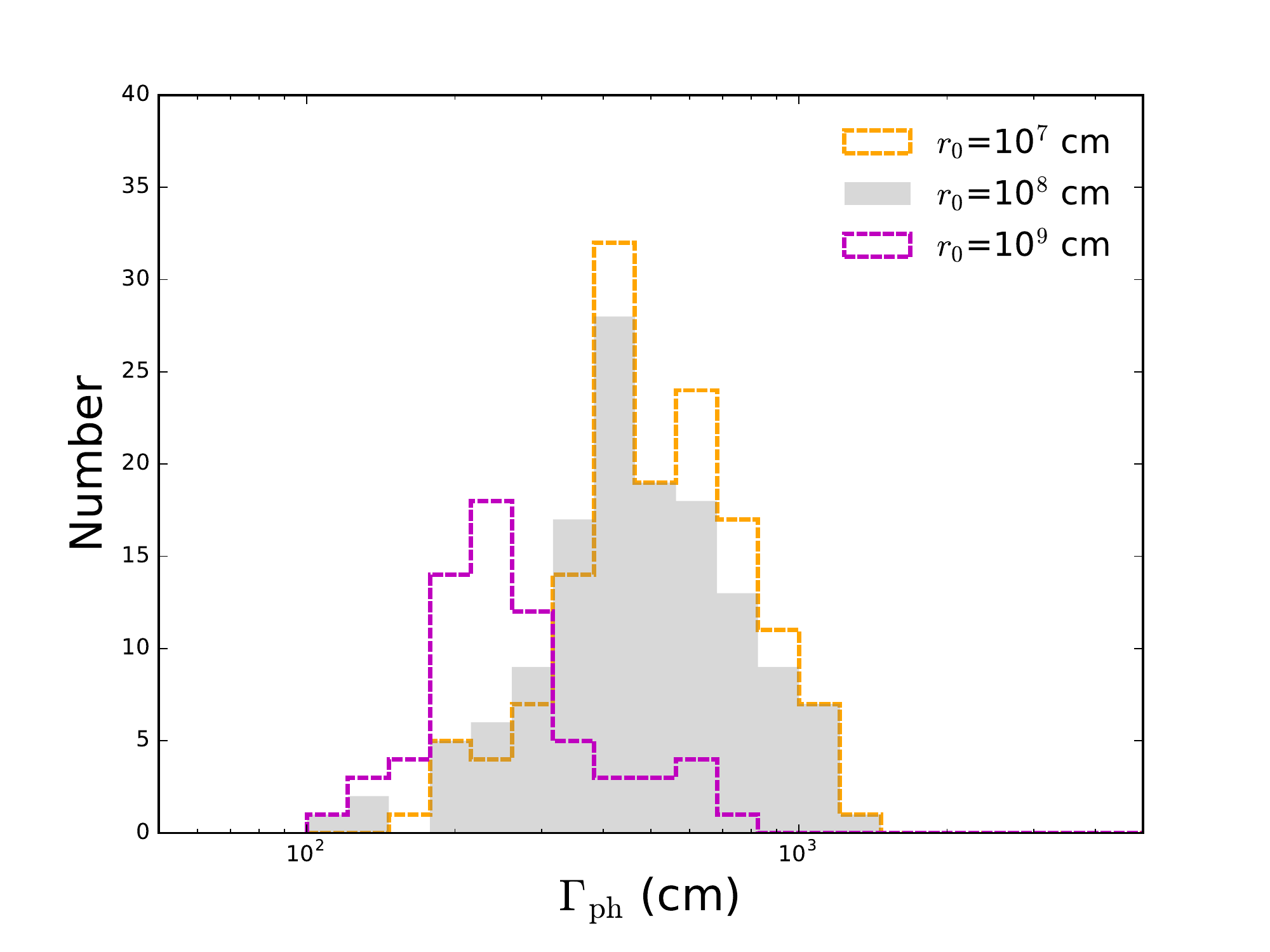}\centering\\
\includegraphics[angle=0, scale=0.45]{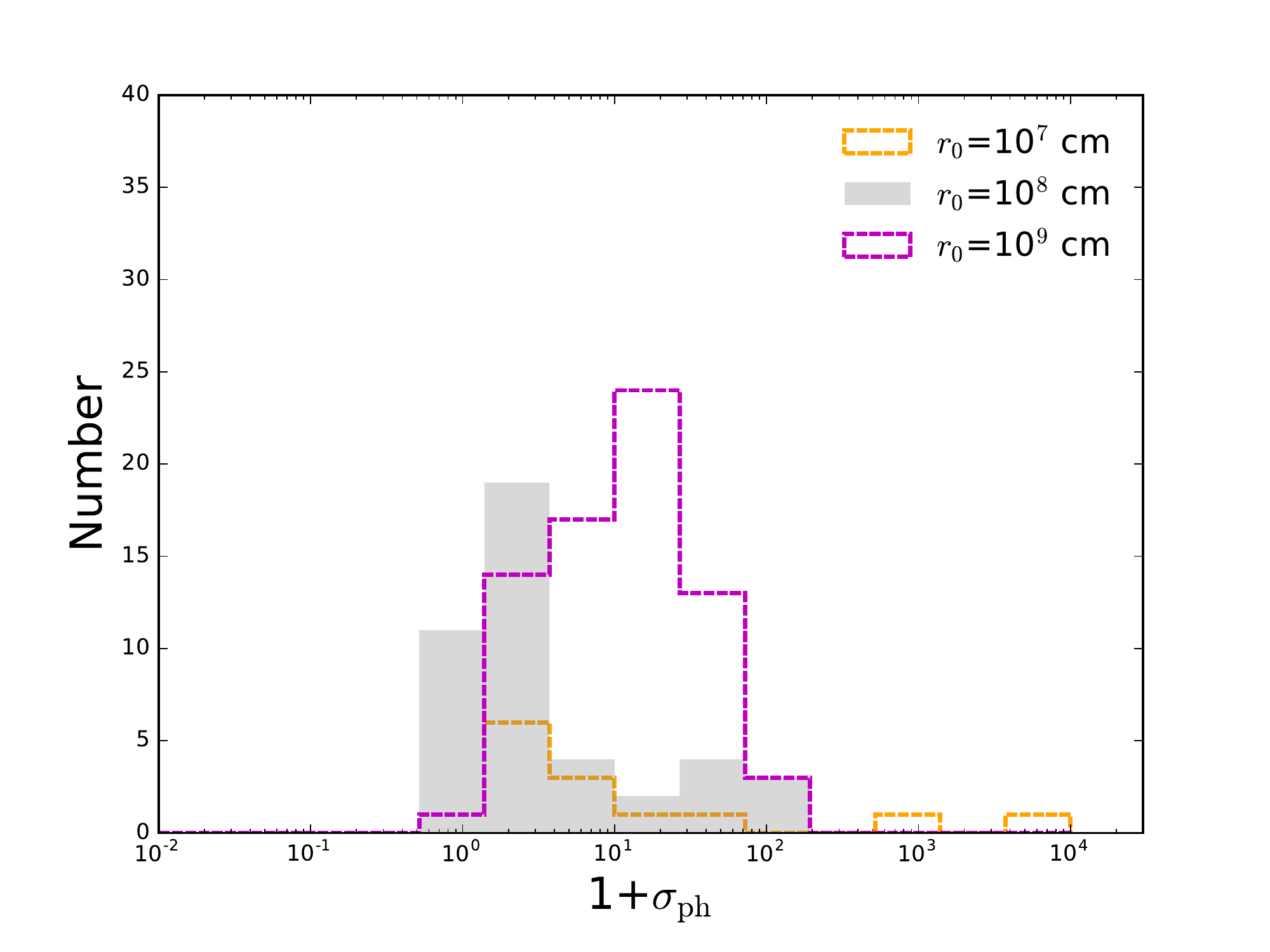}\centering
\includegraphics[angle=0, scale=0.45]{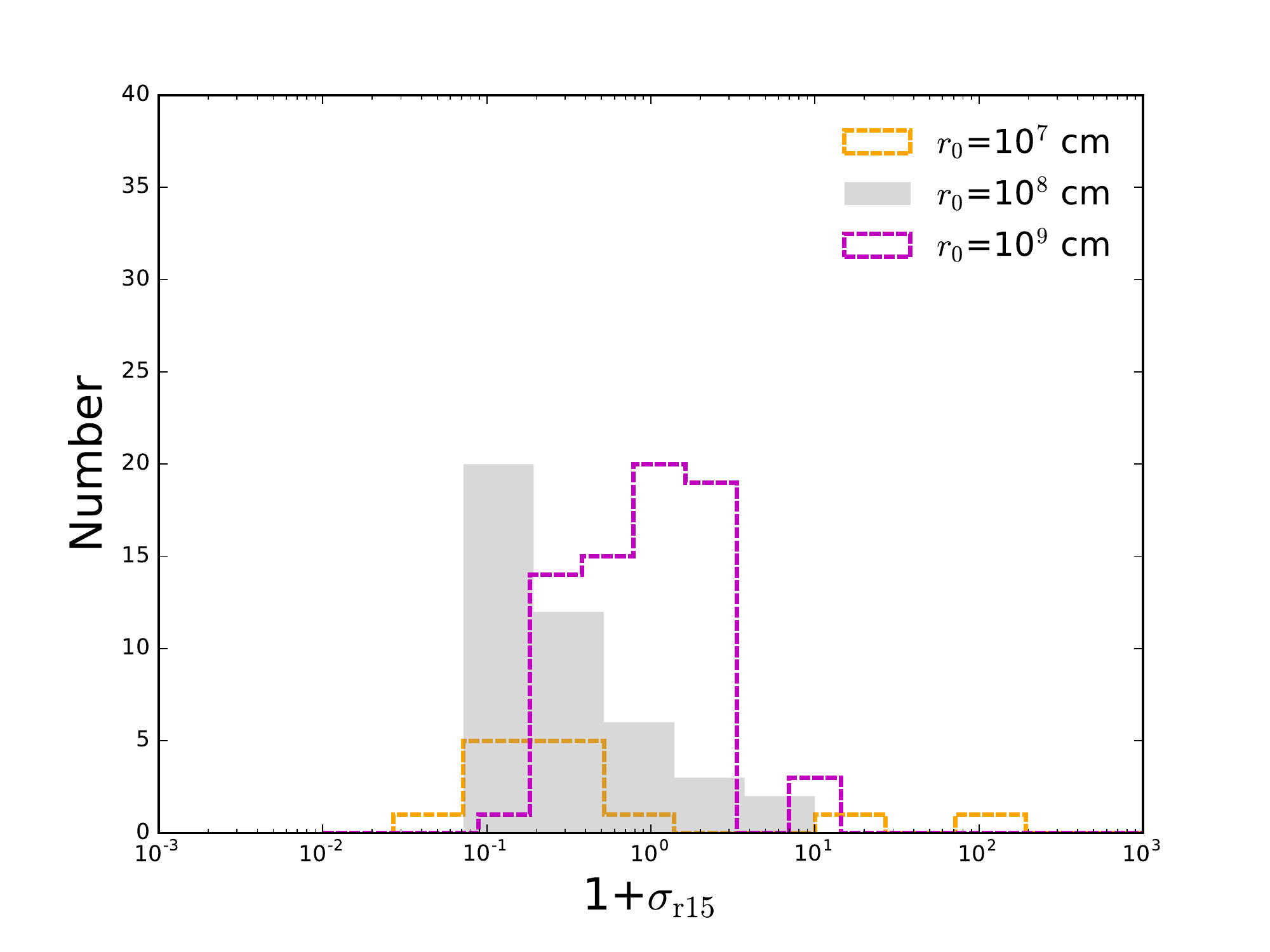}\centering\\
\caption{Distributions of characteristic parameters of the hybrid problem, which is based on the assumptions of constant values of $r_{0}$: $r_{0}$=10$^{7}$ cm (orange), $r_{0}$=10$^{8}$ cm (grey), and $r_{0}$=10$^{9}$ cm (purple). Upper-left panel: for $\eta$-distribution; upper-right panel: for ($1+\sigma_{0}$)-distribution; middle-left panel: for $r_{\rm ph}$-distribution; middle-right panel: for $\Gamma_{\rm ph}$-distribution; bottom-left panel: for ($1+\sigma_{\rm ph}$)-distribution; bottom-right panel: for ($1+\sigma_{\rm r15}$)-distribution.}\label{Distribution}
\end{figure*}

In order to have a global view on the statistical properties of the physical parameters for the hybrid problem, we present the distributions of each relevant physical parameter, comparing them with three typical values of $r_{0}$ (Figure \ref{Distribution}). For a small $r_{0}$, we find $\eta$ tends to be large while (1+$\sigma_{0}$) tends to be small. The peaks of $\eta$ are distributed at $\sim$ 150 for $r_{0}$=10$^{9}$ cm and at $\sim$ 4$\times$10$^{3}$ for $r_{0}$=10$^{7}$ cm, while the peaks of (1+$\sigma_{0}$) are close to unity for $r_{0}$=10$^{7}$ cm and $\sim$ 10 for $r_{0}$=10$^{9}$ cm. We find that (1+$\sigma_{0}$) typically ranges within (1 $\sim$ 100) for all selected $r_{0}$ values. On the other hand, both $r_{\rm ph}$ and $\Gamma_{\rm ph}$ generally share the same peak between different values of $r_{0}$ (except for $r_{0}$=10$^{9}$ cm), in which the peaks are around 10$^{12}$ cm for $r_{\rm ph}$ and $\sim$ 500 for $\Gamma_{\rm ph}$. We do not find a clear trend for (1+$\sigma_{\rm 15}$) and (1+$\sigma_{\rm r15}$) due to a small sample size. 

\begin{figure*}
\includegraphics[angle=0, scale=0.45]{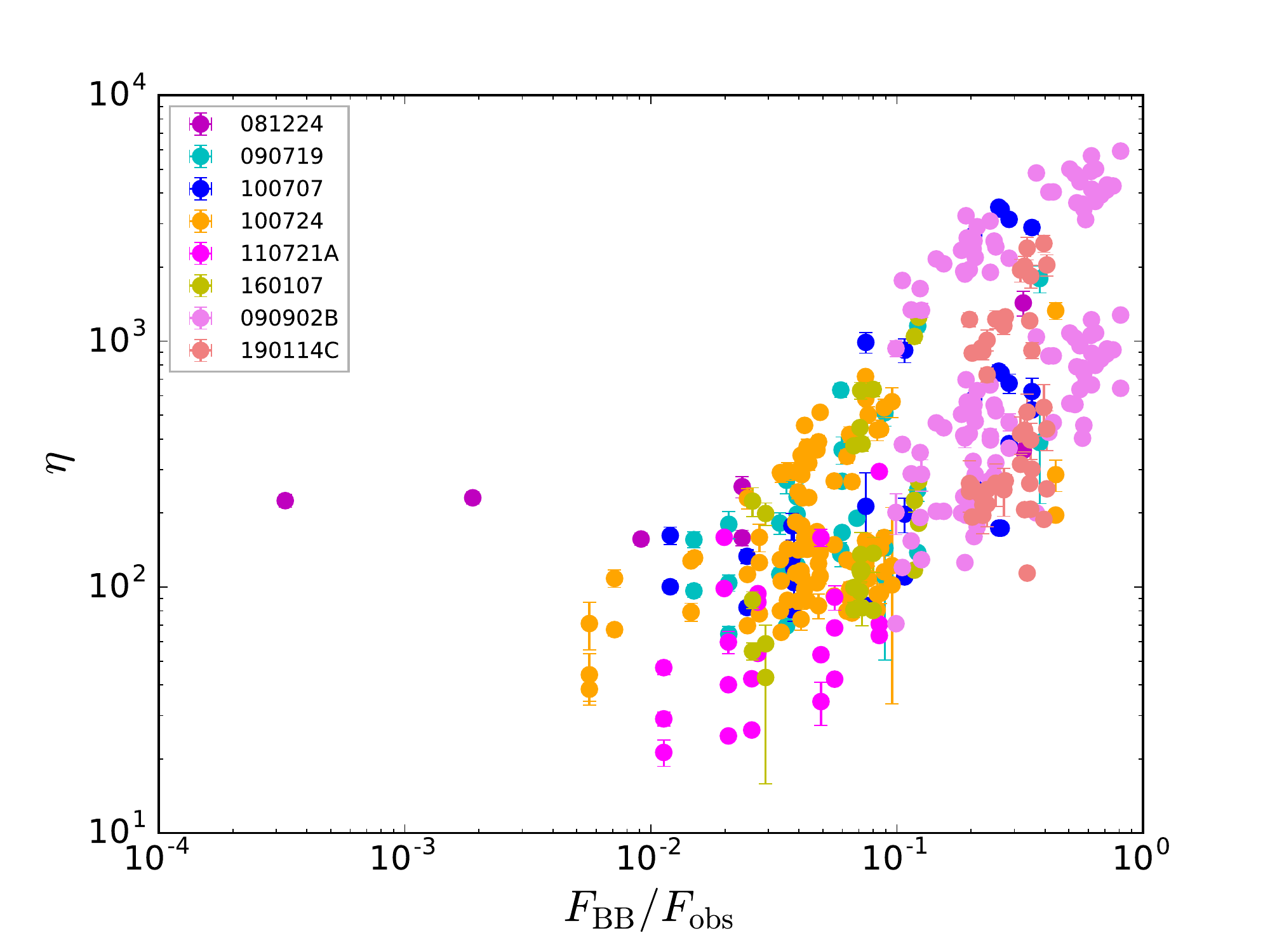}\centering
\includegraphics[angle=0, scale=0.45]{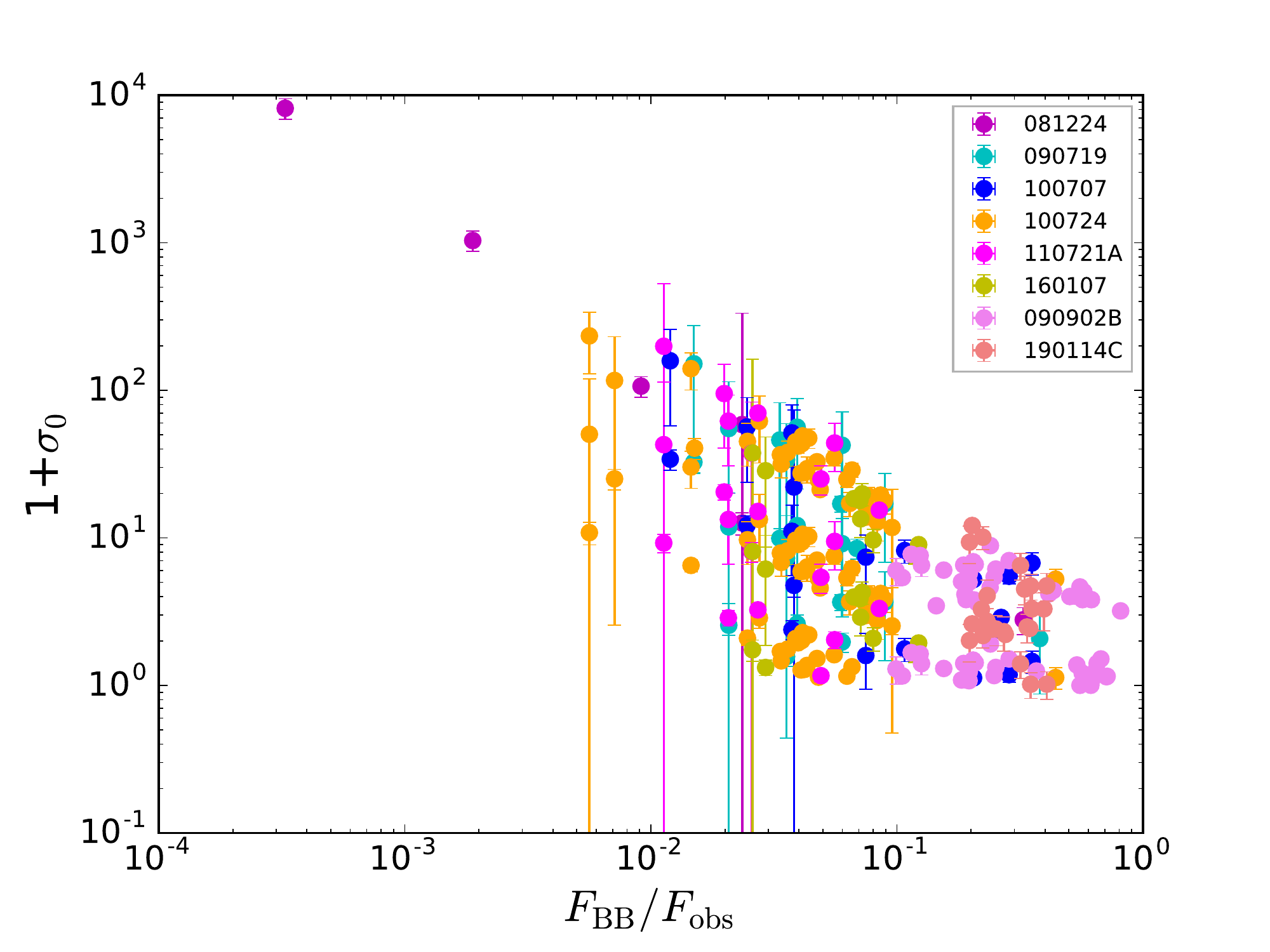}\centering\\
\includegraphics[angle=0, scale=0.45]{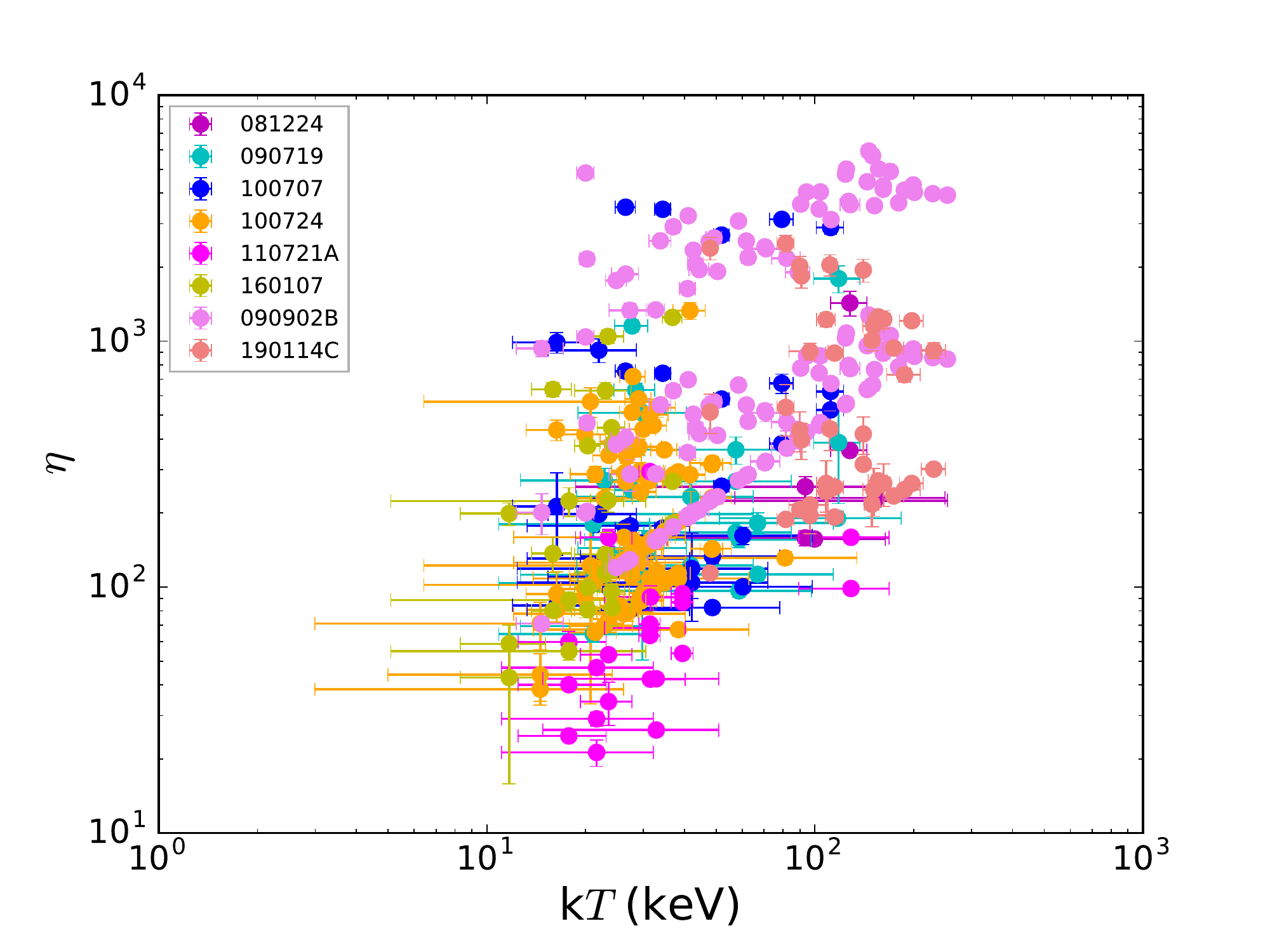}\centering
\includegraphics[angle=0, scale=0.45]{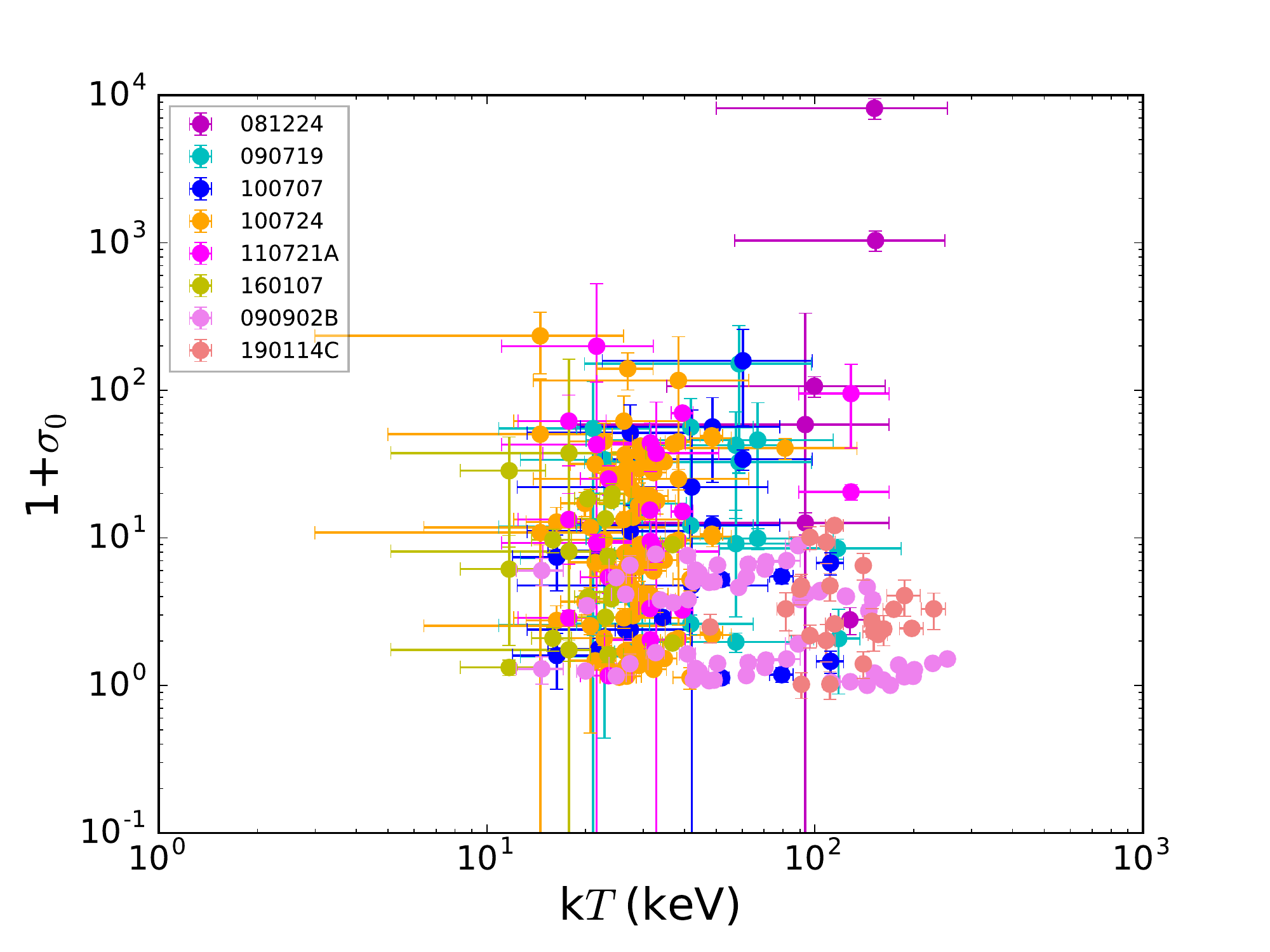}\centering\\
\includegraphics[angle=0, scale=0.45]{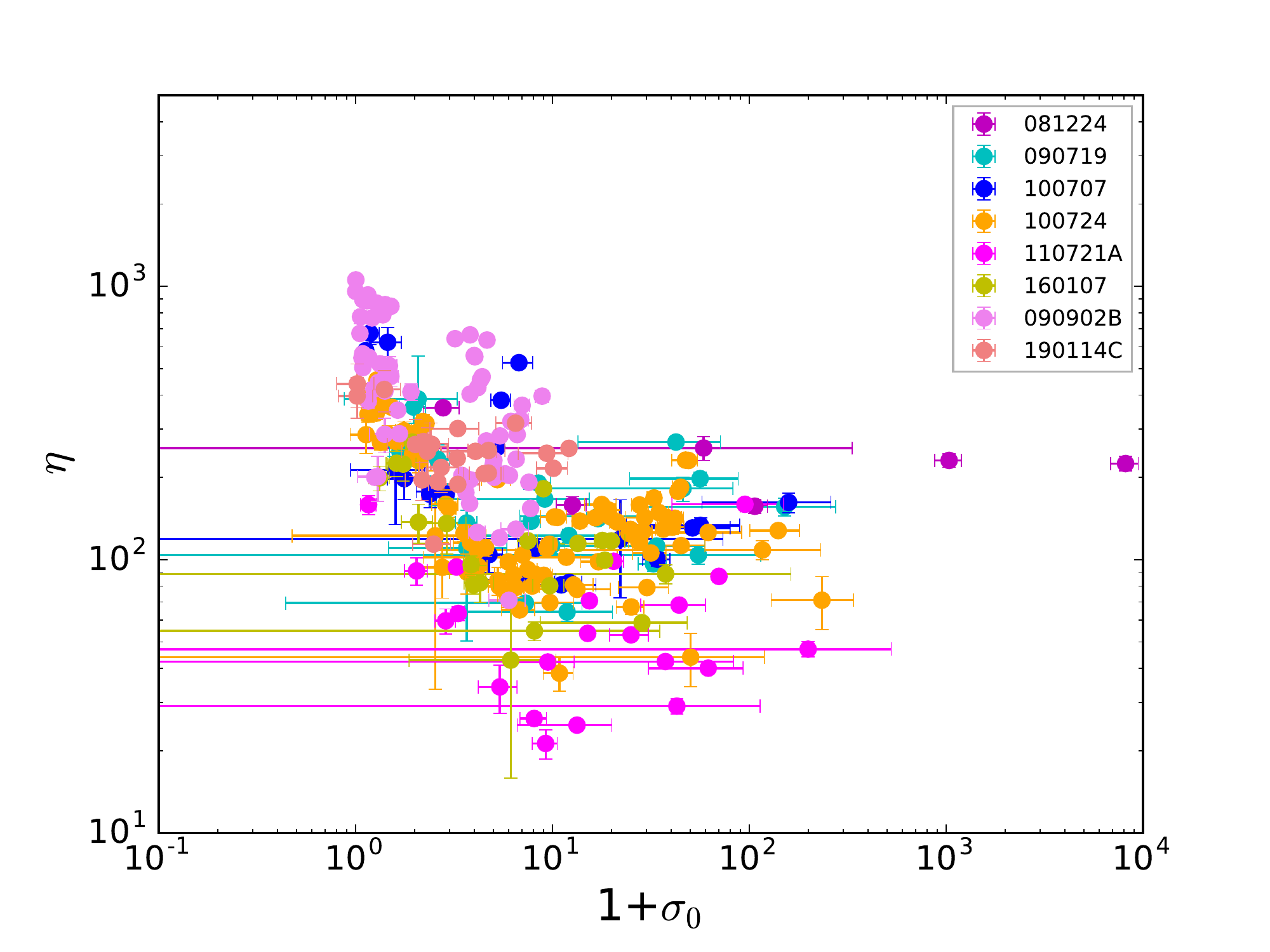}\centering
\includegraphics[angle=0, scale=0.45]{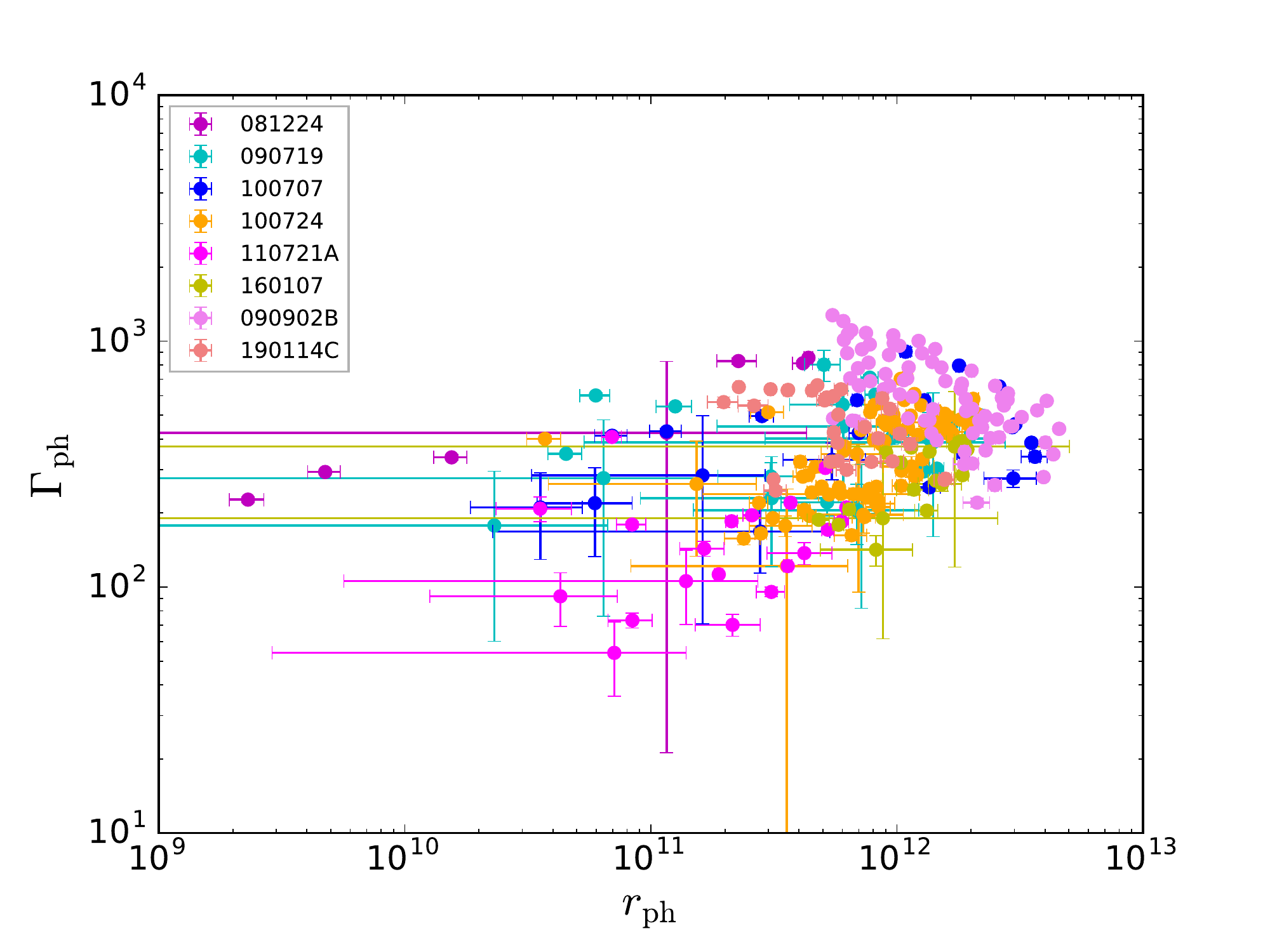}\centering
\caption{Scatter plots of several characteristic parameters of the hybrid problem (based on all $r_{0}$): the $\eta$-($F_{\rm BB}/F_{\rm obs}$) plot (upper-left panel), the ($1+\sigma_{0}$)-($F_{\rm BB}/F_{\rm obs}$) plot (upper-right panel), the $\eta$-$kT$ plot (middle-left panel); the ($1+\sigma_{0}$)-$kT$ plot (middle-right panel), the $\eta$-($1+\sigma_{0}$) plot (bottom-left panel), and the $\Gamma_{\rm ph}$-$r_{\rm ph}$ plot (bottom-right panel).}\label{Correlation}
\end{figure*}

In Figure \ref{Correlation}, we display some key correlation analysis for the hybrid parameters. We find that the $\eta$-($F_{\rm BB}/F_{\rm obs}$) plot shows a clear monotonous-positive relation, whereas both the (1+$\sigma_{0}$)-($F_{\rm BB}/F_{\rm obs}$) and $\eta$-(1+$\sigma_{0}$) plots present a strong monotonous-negative relation. The results are consistent with the predicted expectation in the hybrid model---a high thermal flux ratio tends to be a high $\eta$ and small (1+$\sigma_{0}$). The thermal flux ratio and $\eta$ track each other since both denote the strength of the thermal component. Therefore, both of them have an opposite relation with (1+$\sigma_{0}$). For $r_{\rm ph}$-$\Gamma_{\rm ph}$, we also find a positive relation. For $\eta$-$kT$ and (1+$\sigma_{0}$)-$kT$ relations, we do not find a clear trend.

\section{Conclusion and Discussion}\label{Conclusion}

GRB jets are more likely to originate from a hybrid system, which consists of a quasi-thermal (hot fireball) component as well as a nonthermal (cold Poynting-flux) component at the central engine concomitantly. The hybrid model has been discussed in detail in \cite{2015ApJ...801..103G}. However, it has not yet been applied to a large sample of {\it Fermi} GRBs. In this paper, we first applied the top-down approach of \cite{2015ApJ...801..103G} to diagnose a large sample of {\it Fermi} GRBs with the detected photosphere component, and then carried out a statistical analysis of the central engine properties. In total, we included eight such GRBs for our analysis (see our Paper I for details). In order to obtain the observational parameters, we first employed a Bayesian analysis and MCMC method to fit our sample. Three observed quantities are obtained, including: BB temperature $kT$, BB flux $F_{\rm BB}$, and thermal flux ratio $F_{\rm BB}/F_{\rm obs}$. After the regime judgment, we inferred all the relevant physical parameters for the hybrid problem from the corresponding formula of each regime (see Appendix \ref{sec:formula}), including $\eta$, (1+$\sigma_{0}$), $r_{\rm ph}$, $\Gamma_{\rm ph}$, (1+$\sigma_{\rm ph}$), and (1+$\sigma_{\rm r15}$). Our analysis is based on the assumption that $r_{0}$ is a constant. Considering several realistic scenarios for a central engine, we adopted three typical values of $r_{0}$: $r_{0}$=10$^{7}$cm, $r_{0}$=10$^{8}$cm, and $r_{0}$=10$^{9}$cm. For the busts without redshift observation, we use a typical value ($z$ = 2) instead. After analyzing the evolutionary properties of the physical parameters in our sample, we found $\eta \gg 1$ in all time bins of all bursts, indicating a hot fireball component. We also found that in some time bins in five bursts (GRB 081224, GRB 110721A, GRB 090719, GRB 100707, and GRB 100724) the derived (1+$\sigma_{0}$) is greater than unity for all selected $r_{0}$ values, implying that a cold Poynting-flux component may also play an important role for these GRBs, and therefore the hybrid jet problem must be involved. The other three bursts (GRB 190114C, GRB 090902B, and GRB 160107A) show $r_{0}$-dependent behavior, which means whether this is possible or not depends on which $r_{0}$ is the true size at the central engine. If $r_{0}$ is small (=10$^{7}$cm), one has (1+$\sigma_{0}$) $\sim$ 1, this in agreement with the case of $\eta$$\gg$1 and $\sigma_{0}$$\ll$1 in the hybrid problem; if $r_{0}$ is large (=10$^{9}$cm), we have (1+$\sigma_{0}$) $>$ 1, this still can be explained by the hybrid problem, where $\eta$ is smaller and $\sigma_{0}$ is larger. Interestingly, we found that (1+$\sigma_{\rm r15}$)$>$1 for some time bins for all $r_{0}$ in GRB 081224 and GRB 110721A. This indicates that the ICMART event rather than IS is the mechanism to power the nonthermal emission. Other GRBs, either exhibit $r_{0}$-dependent behavior (GRB 090719, GRB 100707, GRB 100724B, and GRB 160107A), or have no time bin (GRB 190114C and GRB 090902B) satisfying (1+$\sigma_{\rm r15}$) $>$1. Temporal properties of the physical parameter show that basically, the thermal flux ratio is directly proportional to $\eta$, but inversely proportional to (1+$\sigma_{0}$), which is the natural expectation predicted by the hybrid problem. Since a high thermal flux ratio indicates a strong thermal component and a weak cold Poynting-flux component, $\eta$ should be large and (1+$\sigma_{0}$) should be small. Moreover, the global parameter relations show that the $\eta$-($F_{\rm BB}/F_{\rm obs}$) plot presents a monotonic-positive relation, whereas the (1+$\sigma_{0}$)-($F_{\rm BB}/F_{\rm obs}$) plot shows a monotonic-negative relation. In conclusion, in a more general hybrid jet model, which introduces another magnetization parameter $\sigma_{0}$ on the basis of the traditional fireball model, at least a majority of {\it Fermi} bursts (probably all) can be well interpreted.  

Finally, in our analysis, several caveats are worth mentioning. The first one is the problem of the selection value of $r_{0}$. In the hybrid problem, our analysis is based on the assumption of a constant $r_{0}$. We adopted the values of three $r_{0}$ ($r_{0}$=10$^{7}$cm, $r_{0}$=10$^{8}$cm, and $r_{0}$=10$^{9}$cm), which span two orders of magnitude. However, our results significantly vary with different $r_{0}$ values. Since it is impossible to give an accurately true value of $r_{0}$, this leads us to make some not very confident explanations in some cases. For instance, in GRB 09092B, the burst has the highest thermal flux ratio. When $r_{0}$=10$^{7}$cm, all time bins have (1+$\sigma_{0}$)$\sim$ 1. However, when $r_{0}$=10$^{8}$cm, only a part of time bins show (1+$\sigma_{0}$)$\sim$ 1. Moreover, when $r_{0}$=10$^{9}$cm, no time bin has (1+$\sigma_{0}$)$\sim$ 1; rather, all time bins have (1+$\sigma_{0}$)$>$1. Such $r_{0}$-dependent behavior is evidenced from another burst, GRB 190114C, which also has a very high thermal flux ratio. \cite{2015ApJ...801..103G} studied a case (GRB 110721A) but applied different $r_{0}$ values: $r_{0}$= 10$^{8}$ cm, $r_{0}$= 10$^{9}$ cm, and $r_{0}$= 10$^{10}$ cm. Using $r_{0}$=10$^{10}$ cm for the analysis may be a little big, since the size of a naked engine (a hyper-accreting black hole or a millisecond magnetar) is $r_{0}$ $\sim$ 10$^{7}$ cm, or for a `re-born' fireball (considering an extended envelope of a collapsar progenitor), $r_{0} \sim R_{*} \theta_{j} \sim  10 ^{9} R_{*,10} \theta_{j,-1}$~cm (where $R_{*}$ is the size of the progenitor star, and $\theta_{j}$ is the jet opening angle). On the other hand, only a very small number of time bins of $r_{0}=10^{10}$~cm can go through the regime judgment (one for GRB 100707A, two for GRB 110721A, three for GRB 090902B, and no time bin for other five bursts).

The second one is the redshift problem. In our sample, the redshift of more than half of the GRBs is unknown. However, in reality, the derivation of some physical parameters require a redshift measurement. In order to test the effect of various redshift values on the results, we compare the temporal properties of the physical parameters with five different $z$ values for GRB 110721A (Figure \ref{redshift}): $z$ = 0.382, $z$ = 1, $z$ = 2, $z$ = 3.512, and $z$ = 8. For simplicity, our test is only based on a typical radius, $r_{0}$ = 10$^{8}$ cm. 0.382 and 3.512 are two candidates of observed values of redshift for GRB 110721A, as reported in \cite{2011GCN.12193....1B}, and the former is preferred. We find that the effect of redshift is moderate, which has an impact on the results within one order of magnitude. Therefore, our calculations are adopted a typical value ($z$ = 2) for the bursts, whose redshift is unknown. More interestingly, we find that $\eta$, $r_{\rm ph}$, and $\Gamma_{\rm ph}$ are more sensitive than (1+$\sigma_{0}$), (1+$\sigma_{\rm ph}$), and (1+$\sigma_{\rm r15}$) replying on the selection of redshift. However, for those GRBs without redshift measurement, we still need to be cautious in explaining the physical parameters.

Last, our current work is based on the assumption that GRBs have a jet structure. There are some other models (e.g., Induced Gravitational Collapse model) may also well account for the observations. For example, in recent months there has been the identification of the GRB ``inner engine" in GRB 130427A \citep{2019ApJ...886...82R}. This inner engine, applied also to GRB 190114C, GRB 160509A and GRB 160625B \citep{2019arXiv191012615L} evidenced that the MeV radiation observed by {\it Fermi}-GBM occurs close to the Black Hole, is not collimated and has a self-similar temporal structure. Quantized GeV emission, observed by {\it Fermi} Large Area Telescope, originates very close to the Black Hole horizon and represents the GRB jetted emission \citep{2020EPJC...80..300R}.

\acknowledgments

I appreciate the valuable comments from the anonymous referee. I would like to thank Dr. Yu Wang, Prof. Felix Ryde, Prof. He Gao,  Dr. Rahim Moradi, Prof. She Sheng Xue, Prof. Gregory Vereshchagin, Prof. Jorge Rueda, and Prof. Carlo Luciano Bianco for useful discussions, and the support from Prof. Remo Ruffini. I particularly thank to Prof. Bing Zhang for giving an initial idea on the subject and much useful discussions. This research made use of the High Energy Astrophysics Science Archive Research Center (HEASARC) Online Service at the NASA/Goddard Space Flight Center (GSFC). 

%\clearpage
\vspace{5mm}
\facilities{{\it Fermi}/GBM}
\software{3ML \citep{2015arXiv150708343V}, MULTINEST \citep{2008MNRAS.384..449F, 2009MNRAS.398.1601F, 2019OJAp....2E..10F}, emcee \citep{2013PASP..125..306F}, and Matplotlib \citep{2007CSE.....9...90H}}
\bibliography{../../MyBibFiles/MyBibFile.bib}

\begin{deluxetable}{cccccccccc}
%\rotate
%\tablewidth{0pt}
\tabletypesize{\scriptsize}
\tablecaption{Photosphere Properties of the Hybrid Jet Problem of Our Sample.}
\tablenum{1}
\tablehead{
\colhead{GRB}
&\colhead{$z$}
&\colhead{$r_{\rm 0}$}
&\colhead{Spectrum\tablenotemark{a} (Overall)}
&\colhead{$\eta$}
&\colhead{(1+$\sigma_{0}$)}
&\colhead{$r_{\rm ph}$}
&\colhead{$\Gamma_{\rm ph}$}
&\colhead{(1+$\sigma_{\rm ph}$)}
&\colhead{(1+$\sigma_{\rm r15}$)}\\
&\colhead{(Used Value)}
&\colhead{(cm)}
&\colhead{(Number)}
&\colhead{(Evolution)}
&\colhead{(Evolution,$>$1)}
&\colhead{(Evolution)}
&\colhead{(Evolution)}
&\colhead{($>$1)}
&\colhead{($>$1)}
}
\colnumbers
\startdata
081224&2&$10^{7}$&5(5)&{\it d.}&{\it i.}-to-{\it d.}, (4)&{\it d.}-to-{\it i.}&{\it d.}-to-{\it i.}&4&3\\
&&$10^{8}$&2(5)&{\it d.}&{\it i.}, (2)&{\it d.}&{\it d.}&1&1\\
&&$10^{9}$&0(5)&...&...&...&...&...&...\\
\hline
090719&2&$10^{7}$&11(12)&{\it d.}&{\it i.}, (7)&{\it f.}-to-{\it d.}&{\it f.}-to-{\it d.}&3&0\\
&&$10^{8}$&10(12)&{\it d.}&{\it i.}, (10)&{\it f.}-to-{\it d.}&{\it f.}-to-{\it d.}&6&2\\
&&$10^{9}$&7(12)&{\it d.}&{\it i.}, (7)&{\it f.}&{\it f.}&7&4\\
\hline
100707&2&$10^{7}$&11(11)&{\it f.}-to-{\it d.}&{\it f.}-to-{\it i.}, (4)&{\it d.}&{\it i.}&2&0\\
&&$10^{8}$&11(11)&{\it f.}-to-{\it d.}&{\it d.}-to-{\it i.}, (9)&{\it i.}-to-{\it d.}&{\it f.}-to-{\it d.}&4&2\\
&&$10^{9}$&8(11)&{\it d.}&{\it d.}-to-{\it i.}, (8)&{\it i.}-to-{\it d.}&{\it f.}-to-{\it d.}&7&1\\
\hline
100724B&2&$10^{7}$&32(33)&{\it i.}&{\it d.}, (22)&{\it f.}&{\it f.}&2&0\\
&&$10^{8}$&31(33)&{\it f.}&{\it d.}, (31)&{\it f.}&{\it f.}&19&1\\
&&$10^{9}$&30(33)&{\it f.}&{\it d.}, (30)&{\it f.}&{\it f.}&30&23\\
\hline
110721A&0.382&$10^{7}$&8(10)&{\it d.}&{\it d.}-to-{\it i.}, (7)&{\it i.}-to-{\it d.}&{\it d.}&1&0\\
&&$10^{8}$&8(10)&{\it d.}&{\it d.}-to-{\it i.}, (8)&{\it i.}-to-{\it d.}&{\it d.}&6&1\\
&&$10^{9}$&6(10)&{\it d.}&{\it d.}-to-{\it i.}, (6)&{\it i.}-to-{\it d.}&{\it d.}&6&4\\
\hline
190114C&0.424&$10^{7}$&17(18)&{\it d.}&{\it d.}, (0)&{\it i.}&{\it d.}&0&0\\
&&$10^{8}$&17(18)&{\it f.}&{\it d.}, (12)&{\it i.}&{\it f.}-to-{\it d.}&3&0\\
&&$10^{9}$&9(18) &{\it f.}&{\it f.}, (9)&{\it i.}&{\it f.}-to-{d.}&8&0\\
\hline
090902B&1.882&$10^{7}$&47(48)&{\it f.}-to-{\it d.}&{\it f.}, (0)&{\it f.}-to-{\it d.}&{\it f.}-to-{\it d.}&0&0\\
&&$10^{8}$&47(48)&{\it f.}-to-{\it d.}&{\it f.}, (29)&{\it f.}-to-{\it d.}&{\it f.}-to-{\it d.}&0&0\\
&&$10^{9}$&33(48)&{\it f.}-to-{\it d.}&{\it f.}, (33)&{\it f.}-to-{\it i.}&{\it f.}-to-{\it d.}&31&0\\
\hline
160107A&2&$10^{7}$&9(9)&{\it f.}&{\it f.}, (2)&{\it f.}&{\it f.}&0&0\\
&&$10^{8}$&9(9)&{\it f.}&{\it f.}, (9)&{\it f.}&{\it f.}&1&0\\
&&$10^{9}$&9(9)&{\it f.}&{\it f.}, (9)&{\it f.}&{\it f.}&9&2\\
\enddata\label{tab:1}
\vspace{3mm}
Note.The parameters we list include: GRB name (Column 1), used value of redshift (Column 2), used value of $r_{0}$ (Column 3), time bin of passed regime judgements and total (Column 4), temporal properties of $\eta \gg$1 (Column 5), time bin of (1+$\sigma_{0}$)$>$1 (Column 6), temporal properties of $r_{\rm ph}$ (Column 7) and $\Gamma_{\rm ph}$ (Column 8), time bin of (1+$\sigma_{\rm ph}$)$>$1 (Column 9), and time bin of (1+$\sigma_{\rm r15}$)$>$1 (Column 10).
\tablenotetext{a}{The inferred physical parameters are based on different regimes defined for the hybrid problem, which requires regime judgment, see Table 2 of \cite{2015ApJ...801..103G}. To ensure that our methods are correct, we first adopt the same spectral data \citep[obtained from][]{2013MNRAS.433.2739I} and values of $r_{0}$ ($r_{0}$=10$^{8}$ cm, $r_{0}$=10$^{9}$ cm, and $r_{0}$=10$^{10}$) for a test case (GRB 110721A) as also used in \cite{2015ApJ...801..103G}. We find our results are the same as that of \cite{2015ApJ...801..103G}, indicating our approaches are correct.}
\end{deluxetable}

\clearpage

\clearpage

\clearpage

\clearpage

\clearpage

\clearpage
\appendix

\setcounter{figure}{0}    
\setcounter{section}{0}
\setcounter{table}{0}
\renewcommand{\thesection}{A\arabic{section}}
\renewcommand{\thefigure}{A\arabic{figure}}
\renewcommand{\thetable}{A\arabic{table}}
\renewcommand{\theequation}{A\arabic{equation}}

In this appendix, we present the definition of used models (Appendix \ref{sec:model}); the expressions of the regimes II, III, VI and V of `top-down' approach (Appendix \ref{sec:formula}); and provide additional figures to show the temporal evolution of the physical parameters of the hybrid problem for GRB 110721A but based on the fitted parameters obtained from \cite{2013MNRAS.433.2739I} with different redshift (Appendix \ref{sec:addfigures}).

\section{Definition of Models \label{sec:model}}

The Band function \citep{1993ApJ...413..281B} in the photon number spectrum is defined as
\begin{eqnarray}
\label{n}  f_{\rm BAND}(E)=A \left\{ \begin{array}{ll}
(\frac{E}{E_{\rm piv}})^{\alpha} \rm exp (-\frac{{\it E}}{{\it E_{0}}}), & E \le (\alpha-\beta)E_{0}  \\
\lbrack\frac{(\alpha-\beta)E_{0}}{E_{\rm piv}}\rbrack^{(\alpha-\beta)} \rm exp(\beta-\alpha)(\frac{{\it E}}{{\it E_{\rm piv}}})^{\beta}, & E\ge (\alpha-\beta)E_{0}\\
\end{array} \right.
\end{eqnarray}
where
\begin{equation}
E_{\rm p}=(2+\alpha)E_{0},
\end{equation}
where $A$ is the normalization factor at 100 keV in units of ph cm$^{-2}$keV$^{-1}$s$^{-1}$, $E_{\rm piv}$ is the pivot energy fixed at 100 keV, $\alpha$ and $\beta$ are the low-energy and high-energy power-law photon spectral indices, respectively. 
The two spectral regimes are separated by the break energy $E_{0}$ in units of keV, and $E_{\rm p}$ is the peak energy in the $\nu F_{\nu}$ space in units of keV.

The cutoff power law, or the so-called Comptonized model (COMP), which is written as
\begin{equation}
f_{\rm COMP}(E) =A \left(\frac{E}{E_{\rm piv}}\right)^{\alpha}\rm exp(-\frac{\it E}{\it E_{0}})
\label{CPL}
\end{equation}

The single power law is defined as
\begin{equation}
f_{\rm PL}(E) =A \left(\frac{E}{E_{\rm piv}}\right)^{\Gamma}
\label{CPL}
\end{equation}
where $A$ is the normalization and $\Gamma$ is the spectral index.

The BB emission can be modified by Planck spectrum, which is given by the photon flux
\begin{equation}
f_{\rm BB}(E,t)=A(t)\frac{E^{2}}{\rm exp\lbrack \frac{{\it E}}{{\it kT}(t)}\rbrack-1},
\end{equation}
where $E$ is the photon energy, $k$ is the Boltzmann constant. 
The BB emission depends on two free parameters only: temperature, $T(t)$, and the normalization, $A(t)$.

\clearpage
\section{Formalism of `Top-down' Approach \label{sec:formula}}
For regime II (see also Eq.(37) in \citealt{2015ApJ...801..103G}), we have:
\begin{align}
\begin{split}
& 1+\sigma_{0}=25.5(1+z)^{4/3} \left(\frac{kT_{\rm obs}}{50 \rm keV }\right)^{4/3} \times \left(\frac{F_{\rm BB}}{10^{-8} \rm erg s^{-1} cm^{-2}}\right)^{-1/3}r^{2/3}_{0,9}f^{-1}_{\rm th,-1}f^{-1}_{\gamma}d^{-2/3}_{L,28},\\
& \eta =74.8(1+z)^{11/12} \left(\frac{kT_{\rm obs}}{50 \rm keV }\right)^{11/12} \times \left(\frac{F_{\rm BB}}{10^{-8} \rm erg s^{-1} cm^{-2}}\right)^{1/48}r^{5/24}_{0,9}d^{1/24}_{L,28},\\
& r_{\rm ph} =1.78\times10^{10}{\rm cm}(1+z)^{-25/12} \left(\frac{kT_{\rm obs}}{50 \rm keV }\right)^{-25/12} \times \left(\frac{F_{\rm BB}}{10^{-8} \rm erg s^{-1} cm^{-2}}\right)^{37/48}r^{-7/24}_{0,9}d^{37/24}_{L,28},\\
& \Gamma_{\rm ph} =46.4(1+z)^{-1/12} \left(\frac{kT_{\rm obs}}{50 \rm keV }\right)^{-1/12} \times \left(\frac{F_{\rm BB}}{10^{-8} \rm erg s^{-1} cm^{-2}}\right)^{13/48}r^{-7/24}_{0,9}d^{13/24}_{L,28},\\
& 1+\sigma_{\rm ph}=41.2(1+z)^{7/3} \left(\frac{kT_{\rm obs}}{50 \rm keV }\right)^{-7/12} \times \left(\frac{F_{\rm BB}}{10^{-8} \rm erg s^{-1} cm^{-2}}\right)^{-7/12}r^{7/6}_{0,9}f^{-1}_{\rm th,-1}f^{-1}_{\gamma}d^{-7/6}_{L,28},\\
& 1+\sigma_{r15}=1.08(1+z)^{59/36} \left(\frac{kT_{\rm obs}}{50 \rm keV }\right)^{59/36} \times \left(\frac{F_{\rm BB}}{10^{-8} \rm erg s^{-1} cm^{-2}}\right)^{-47/144}r^{77/72}_{0,9}f^{-1}_{\rm th,-1}f^{-1}_{\gamma}d^{-47/72}_{L,28},\\
\label{eq:regime II}
\end{split}
\end{align}

For regime III and regime VI (see also Eq.(38) in \citealt{2015ApJ...801..103G}), we have:
\begin{align}
\begin{split}
& 1+\sigma_{0}=5.99(1+z)^{4/3} \left(\frac{kT_{\rm obs}}{50 \rm keV }\right)^{4/3} \times \left(\frac{F_{\rm BB}}{10^{-8} \rm erg s^{-1} cm^{-2}}\right)^{-1/3}r^{2/3}_{0,9}f^{-1}_{\rm th,-1}f^{-1}_{\gamma}d^{-2/3}_{L,28},\\
& \eta =20.3(1+z)^{-5/6} \left(\frac{kT_{\rm obs}}{50 \rm keV }\right)^{11/12} \times \left(\frac{F_{\rm BB}}{10^{-8} \rm erg s^{-1} cm^{-2}}\right)^{11/24}r^{-2/3}_{0,9}f^{-3/4}_{\rm th,-1}f^{-3/4}_{\gamma}d^{11/12}_{L,28},\\
& r_{\rm ph} =4.09\times10^{11}{\rm cm}(1+z)^{-3/2} \left(\frac{kT_{\rm obs}}{50 \rm keV }\right)^{-5/8} \times \left(\frac{F_{\rm BB}}{10^{-8} \rm erg s^{-1} cm^{-2}}\right)^{5/8}f^{-1/4}_{\rm th,-1}f^{-1/4}_{\gamma}d^{5/4}_{L,28},\\
& \Gamma_{\rm ph} =121.3(1+z)^{1/2} \left(\frac{kT_{\rm obs}}{50 \rm keV }\right)^{1/2} \times \left(\frac{F_{\rm BB}}{10^{-8} \rm erg s^{-1} cm^{-2}}\right)^{1/8}f^{-1/4}_{\rm th,-1}f^{-1/4}_{\gamma}d^{1/4}_{L,28},\\
\label{eq:regime III}
\end{split}
\end{align}

For regime V (see also Eq.(39) in \citealt{2015ApJ...801..103G}), we have:
\begin{align}
\begin{split}
& 1+\sigma_{0}=6.43(1+z)^{4/3} \left(\frac{kT_{\rm obs}}{50 \rm keV }\right)^{4/3} \times \left(\frac{F_{\rm BB}}{10^{-8} \rm erg s^{-1} cm^{-2}}\right)^{-1/3}r^{2/3}_{0,9}f^{-1}_{\rm th,-1}f^{-1}_{\gamma}d^{-2/3}_{L,28},\\
& \eta =105.0(1+z)^{7/6} \left(\frac{kT_{\rm obs}}{50 \rm keV }\right)^{7/6} \times \left(\frac{F_{\rm BB}}{10^{-8} \rm erg s^{-1} cm^{-2}}\right)^{5/24}r^{1/12}_{0,9}f^{1/2}_{\rm th,-1}f^{1/2}_{\gamma}d^{5/12}_{L,28},\\
& r_{\rm ph} =4.62\times10^{10}{\rm cm}(1+z)^{-13/6} \left(\frac{kT_{\rm obs}}{50 \rm keV }\right)^{-13/6} \times \left(\frac{F_{\rm BB}}{10^{-8} \rm erg s^{-1} cm^{-2}}\right)^{17/24}r^{-1/4}_{0,9}f^{-1/6}_{\rm th,-1}f^{-1/6}_{\gamma}d^{17/12}_{L,28},\\
& \Gamma_{\rm ph} =15.3(1+z)^{-1/6} \left(\frac{kT_{\rm obs}}{50 \rm keV }\right)^{-1/6} \times \left(\frac{F_{\rm BB}}{10^{-8} \rm erg s^{-1} cm^{-2}}\right)^{-5/24}r^{-1/4}_{0,9}f^{-1/6}_{\rm th,-1}f^{-1/6}_{\gamma}d^{5/24}_{L,28},\\
& 1+\sigma_{\rm ph}=44.2(1+z)^{8/3} \left(\frac{kT_{\rm obs}}{50 \rm keV }\right)^{8/3} \times \left(\frac{F_{\rm BB}}{10^{-8} \rm erg s^{-1} cm^{-2}}\right)^{-1/3}r_{0,9}f^{-1/3}_{\rm th,-1}f^{-1/3}_{\gamma}d^{-2/3}_{L,28},\\
& 1+\sigma_{r15}=1.59(1+z)^{35/18} \left(\frac{kT_{\rm obs}}{50 \rm keV }\right)^{35/18} \times \left(\frac{F_{\rm BB}}{10^{-8} \rm erg s^{-1} cm^{-2}}\right)^{-7/72}r^{11/12}_{0,9}f^{-7/18}_{\rm th,-1}f^{-7/18}_{\gamma}d^{-7/36}_{L,28}.\\
\label{eq:regime V}
\end{split}
\end{align}
Here note that regime VI has the identical scalings as regime III.
$f_{\gamma}$ is given by $f_{\gamma}$=$L_{\gamma}/L_{\rm w}$, which connects the total flux $F_{\rm obs}$ to the wind luminosity $L_{\rm w}$; and $f_{\rm th}$=$F_{\rm BB}$/$F_{\rm obs}$, is the thermal flux ratio, which can be directly measured from the data. $f_{\gamma}$ and $r_{0}$ are taken as constants and can be estimated to a typical values (e.g., $f_{\gamma}$=0.5 and $r_{0}$=10$^{8}$ cm.)

%\clearpage
\section{Additional Figures  \label{sec:addfigures}}

\begin{figure*}
\includegraphics[angle=0, scale=0.45]{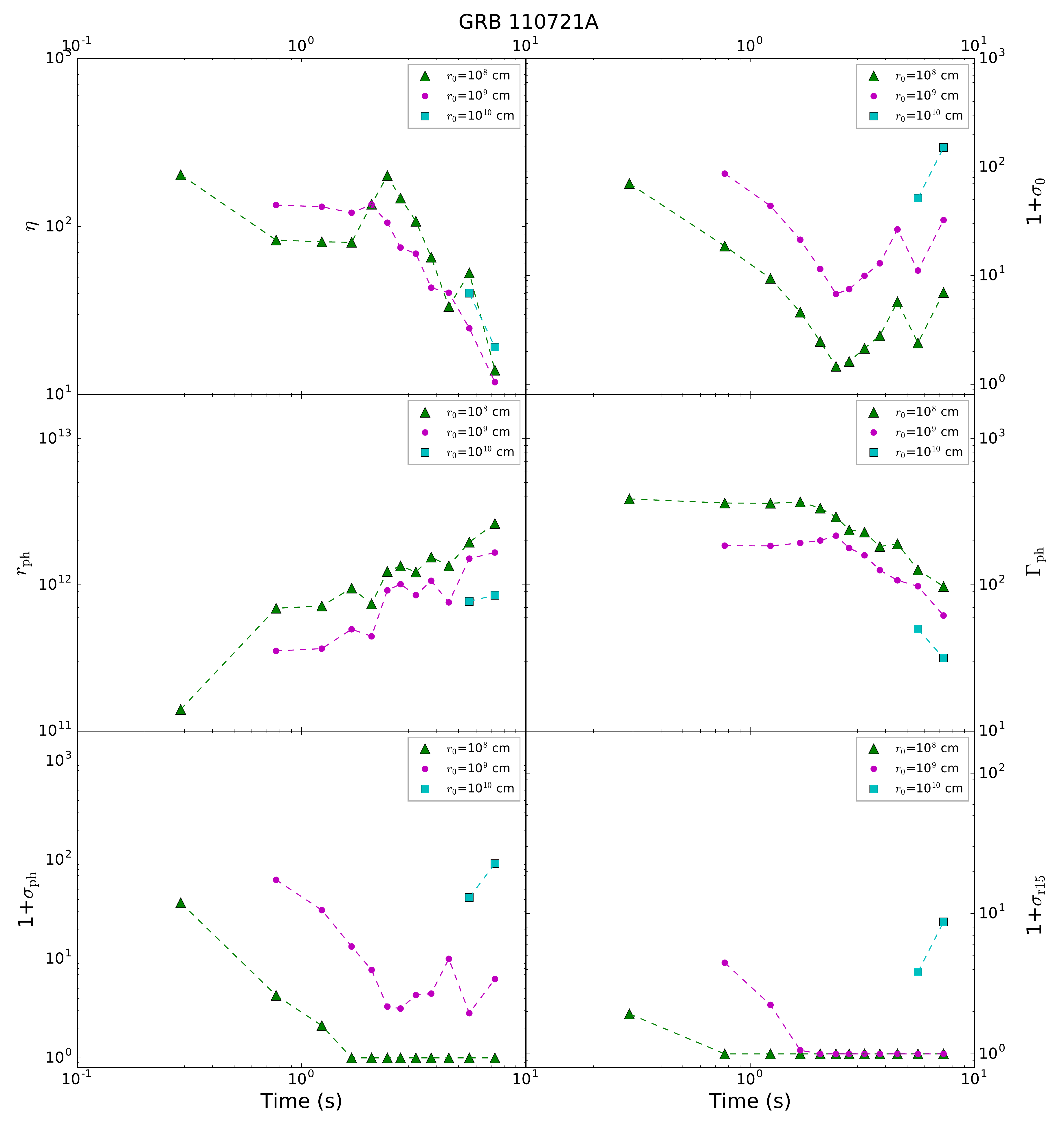}
\caption{Same as Figure \ref{110721200}, but the fitted parameters are adopted from \cite{2013MNRAS.433.2739I}. Here we notice that $\eta$ is less that $\Gamma_{\rm ph}$ in some time bins, which is impossible. The reason is that the jet is still in the acceleration phase; however, we use the coasting phase to derive physical parameters.}\label{110721200A1}
\end{figure*}

\begin{figure*}
\includegraphics[angle=0, scale=0.45]{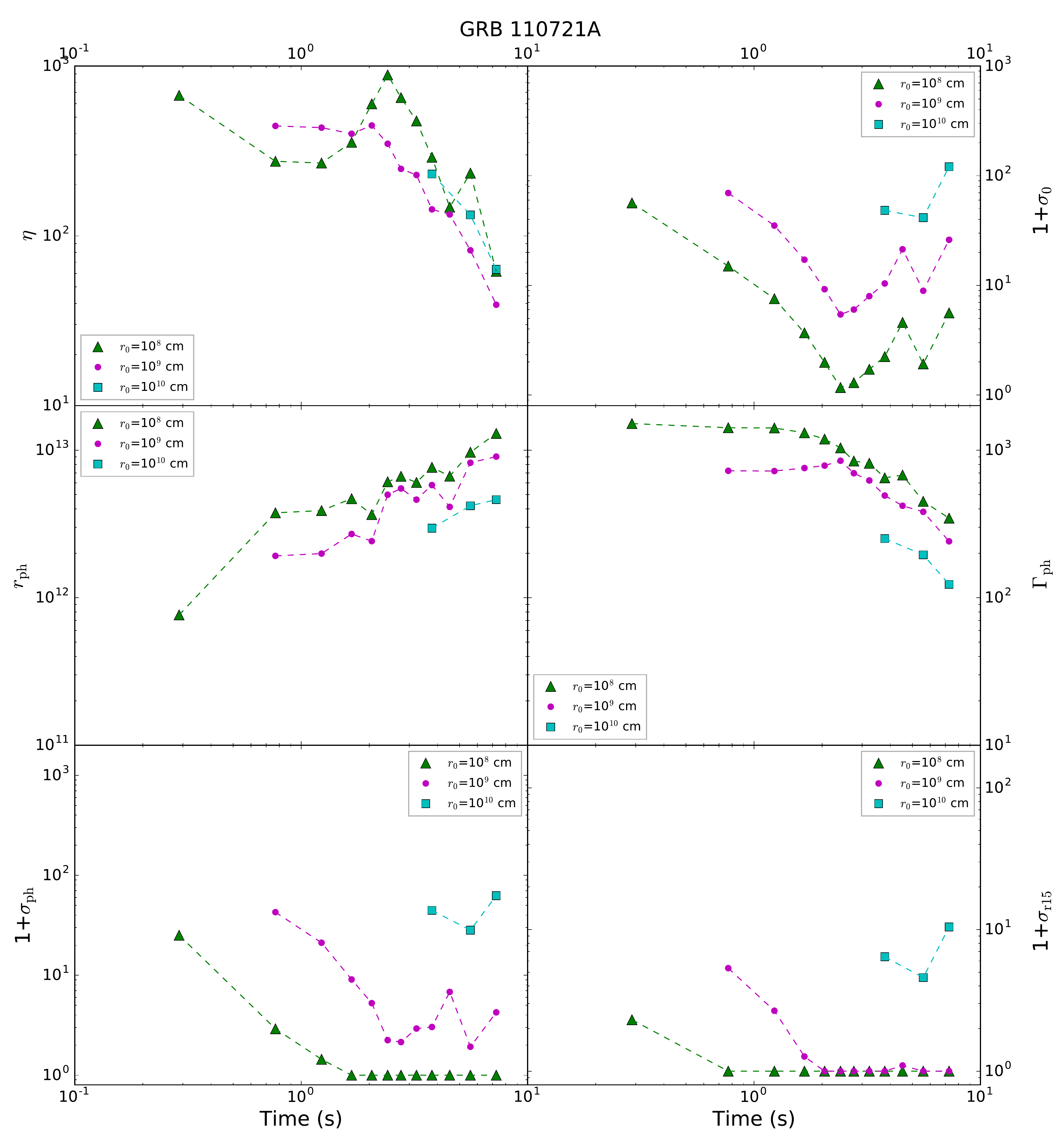}
\caption{Same as Figure \ref{110721200A1}, but redshift is adopted $z$=3.512.}\label{110721200A2}
\end{figure*}

\begin{figure*}
\includegraphics[angle=0, scale=0.45]{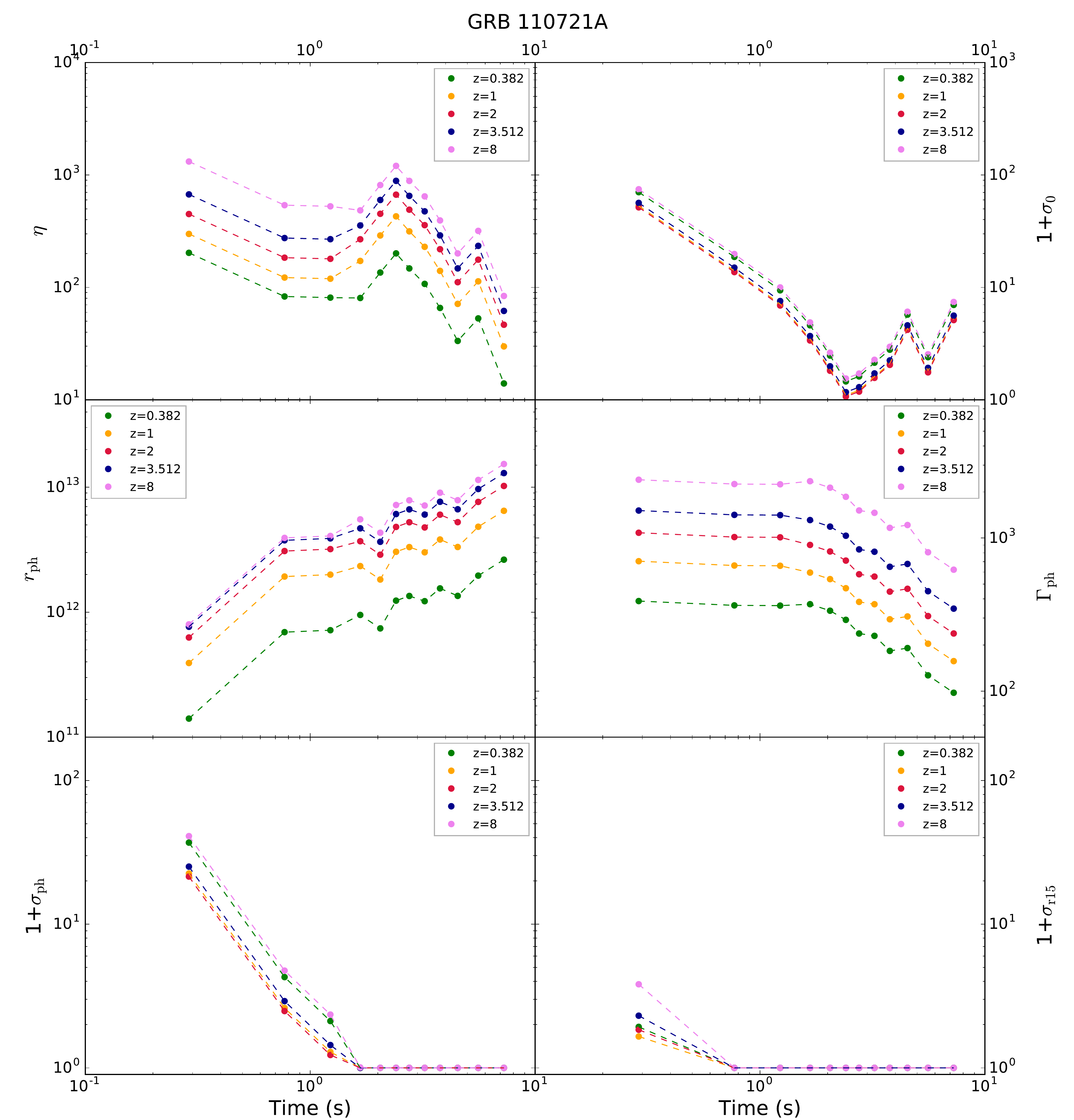}
\caption{Comparison of the evolutional properties of the physical parameters for the hybrid jet problem with different redshifts, which is based on a typical value of $r_{0}$ (=10$^{8}$cm). Different colors indicate different redshift values. The fitted parameters are obtained from \cite{2013MNRAS.433.2739I}.}\label{redshift}
\end{figure*}

Here we show the additional Figure \ref{110721200A1}-\ref{redshift}. For comparison, Figure \ref{110721200A1} and \ref{110721200A2} show the same analysis for a studied case (GRB 110721A) but the fitted parameters are obtained from \cite{2013MNRAS.433.2739I}, which are baed on two candidates of observed values of redshift: $z$=0.382 (Figure \ref{110721200A1}) and $z$=3.512 (Figure \ref{110721200A2}). Figure \ref{redshift} displays the results of the different redshift, which is based on a typical $r_{0}$ value (10$^{8}$ cm). 

\clearpage

\clearpage

\clearpage

\end{document}